\documentclass[prb,showpacs,floatfix,nontitlepage,twocolumn,superscriptaddress]{revtex4-1}
\usepackage[colorlinks=true, citecolor=blue, urlcolor = blue, linkcolor= red, 
bookmarks=true]{hyperref}
\usepackage{natbib}
\usepackage{amsmath}    
\usepackage{graphicx}   
\usepackage{verbatim}   
\usepackage{color}      
\usepackage{subfloat}
\usepackage{hyperref}   
\usepackage{graphicx, psfrag}
\usepackage[centerlast]{caption}
\usepackage{float}
\usepackage[subrefformat=parens,labelformat=parens]{subfig}
\usepackage{siunitx}

\newcommand{\coo}{$\text{CoCr}_2\text{O}_4$}
\newcommand{\tg}{T$_{1g}$}

\begin{document}

\title{Magnons and Magnetodielectric Effects in \coo:  Raman Scattering Studies}

\author{A. Sethi}
\email{asethi8@illinois.edu}
\affiliation{Department of Physics and Frederick Seitz Materials Research Laboratory, University of Illinois, Urbana, Illinois 61801, USA}

\author{T. Byrum}
\affiliation{Department of Physics and Frederick Seitz Materials Research Laboratory, University of Illinois, Urbana, Illinois 61801, USA}

\author{R.D. McAuliffe}
\affiliation{Department of Materials Science and Engineering and \\Frederick Seitz Materials Research Laboratory, University of Illinois, Urbana, Illinois 61801, USA}

\author{S.L. Gleason}
\affiliation{Department of Physics and Frederick Seitz Materials Research Laboratory, University of Illinois, Urbana, Illinois 61801, USA}

\author{J.E. Slimak} 
\affiliation{Department of Physics and Frederick Seitz Materials Research Laboratory, University of Illinois, Urbana, Illinois 61801, USA}

\author{D. P. Shoemaker}
\affiliation{Department of Materials Science and Engineering and \\Frederick Seitz Materials Research Laboratory, University of Illinois, Urbana, Illinois 61801, USA}

\author{S. L. Cooper}
\email{slcooper@illinois.edu}
\affiliation{Department of Physics and Frederick Seitz Materials Research Laboratory, University of Illinois, Urbana, Illinois 61801, USA}

\begin{abstract}
 Magnetoelectric materials have generated wide technological and scientific interest because of the rich phenomena these materials exhibit, including the coexistence of magnetic and ferroelectric orders, magnetodielectric behavior, and exotic hybrid excitations such as electromagnons. The multiferroic spinel material, \coo, is a particularly interesting example of a multiferroic material, because evidence for magnetoelectric behavior in the ferrimagnetic phase seems to conflict with traditional noncollinear-spin-driven mechanisms for inducing a macroscopic polarization. This paper reports an inelastic light scattering study of the magnon and phonon spectrum of \coo\,as simultaneous functions of temperature, pressure, and magnetic field. Below the Curie temperature ($T_C \sim 94$ K) of \coo\,we observe a $\omega \sim 16 \,\text{cm}^{-1}$ $\boldsymbol q=0$ magnon having \tg-symmetry, which has the transformation properties of an axial vector. The anomalously large Raman intensity of the \tg-symmetry magnon is characteristic of materials with a large magneto-optical response and likely arises from large magnetic fluctuations that strongly modulate the dielectric response in \coo. The Raman susceptibility of the \tg-symmetry magnon exhibits a strong magnetic-field dependence that is consistent with the magnetodielectric response observed in \coo, suggesting that magnetodielectric behavior in \coo\,primarily arises from the field-dependent suppression of magnetic fluctuations that are strongly coupled to long-wavelength phonons. Increasing the magnetic anisotropy in \coo\,with applied pressure decreases the magnetic field-dependence of the \tg-symmetry magnon Raman susceptibility in \coo, suggesting that strain can be used to control the magnetodielectric response in \coo.  
\end{abstract}


\maketitle

\section{Introduction}

Multiferroics\textemdash materials exhibiting a coexistence of both magnetic and ferroelectric orders \cite{Kim-2003a,Che-2007}\textemdash have attracted substantial technological and scientific interest recently. The technological interest stems from the multifunctional properties exhibited by multiferroics, which make them potentially useful in device applications such as magnetoelectric memories and switches. Multiferroics are scientifically interesting, in part, because they exhibit a variety of microscopic mechanisms that can result in an interesting interplay between ferroelectric and magnetic orders; \cite{Che-2007} among other consequences, this interplay can spawn interesting dynamical properties in multiferroic materials, including electromagnons, i.e., hybrid excitations involving a coupling between optical phonons and spin waves via the magnetoelectric interaction, \cite{Bar-1970, Pim-2006, Kid-2008a, Kid-2008b, Sus-2008, Val-2009, Ste-2009, Moc-2010, Tiw-2010, Har-2011, Jon-2014, Cao-2015} and magnetodielectric effects. \cite{Kim-2003b, Law-2003, Yan-2012} 

Materials in which geometric frustration leads to non-collinear spin order and strong spin-lattice coupling are particularly rich material environments to find novel magnetoelectric behavior. \cite{Kim-2003a, Got-2004} Transition-metal-oxide spinel materials (\textit{AB}$_2$O$_4$), for example, exhibit both non-collinear spin orders and strong spin-lattice coupling that can lead to magnetoelectric coupling, because the presence of magnetic ions on the \textit{B}-site pyrochlore lattice of the spinel structure often leads to strong geometric frustration and consequent non-collinear orders that can generate multiferroic phenomena. \cite{Che-2007} Magnetoelectric effects are indeed realized in some \textit{A}Cr$_2$O$_4$\, spinels (\textit{e.g.}, \textit{A}=Co$^{2+}$ and Fe$^{2+}$), in which the competition among the various exchange interactions, J$_\text{A-A}$, J$_\text{A-Cr}$, and J$_\text{Cr-Cr}$, involving the \textit{A}$^{2+}$ ions and the Cr$^{3+}$ $S=3/2$ spins lead to complex magnetic orders. \cite{Bor-2009, Sin-2011} 

\coo, in particular, exhibits a succession of magnetic orders, including ferrimagnetic order below $T_C \sim 94$~K, incommensurate conical spiral order below $T_S\sim 26$ K, commensurate order below $T_L \sim 14$ K, \cite{Tom-2004, Tsu-2013} as well as spin-driven multiferroic behavior and dielectric anomalies below $T_S$. \cite{Yam-2006, Law-2006, Cho-2009} Yet, the nature and origin of magnetoelectric behavior in \coo\,remains uncertain. Multiferroicity in \coo\,has been associated with the spin-current mechanism \cite{Kat-2005} involving cycloidal spin order, \cite{Yam-2006} in which the induced electric polarization is generated by the non-collinear spins \cite{Mos-2006} via the inverse Dzyaloshinskii-Moriya interaction, $\boldsymbol P \sim \boldsymbol e_{ij}\times(\boldsymbol S_i\times \boldsymbol S_j)$. However, evidence for multiferroicity, \cite{Sin-2011, Yan-2012} structural distortion, \cite{Yan-2012} and magnetodielectric behavior \cite{Yan-2012} have also been reported above $T_S$ in the ferrimagnetic state of \coo, raising questions about the origin of multiferroic behavior in this material. Yang \textit{et al.}, for example, have suggested that magnetodielectric behavior in \coo\,results from the presence of multiferroic domains that are reoriented in the presence of a magnetic field. \cite{Yan-2012} But magnetodielectric behavior in magnetic materials can also arise from magnetic fluctuations that induce shifts in optical phonon frequencies via strong spin-lattice coupling. \cite{Law-2003}

Unfortunately, a lack of microscopic information regarding spin-lattice coupling has prevented a clear identification of the mechanism for magnetodielectric behavior in \coo. The intersublattice exchange magnon has been observed in \coo\,using infrared and terahertz spectroscopies \cite{Tor-2012, Kam-2013} and optical phonons in \coo\,have been identified using Raman scattering \cite{Kus-2009, Pta-2014, Eft-2015} and optical absorption \cite{Tor-2012} measurements. However, to our knowledge, there have been no microscopic studies of spin-lattice coupling in \coo\,that could clarify the origin of magnetodielectric behavior in this material. The application of pressure \cite{Kan-1988, Tam-1993, Che-2013} would be a useful means of studying spin-lattice coupling and its role in magnetoelectric behavior in spinels such as \coo; indeed, \textit{ab initio} calculations predict that pressure should enhance the macroscopic polarization in the multiferroic regime of \coo. \cite{Eft-2015} However, the effects of pressure on the magnetoelectric behavior and spin-lattice coupling in \coo\,have not yet been experimentally investigated.

Raman scattering is a powerful tool for studying magnons, \cite{Gle-2014, Gim-2016} strong spin-lattice coupling \cite{Gle-2014, Byr-2016} and electromagnons \cite{Caz-2008, Sin-2008, Rov-2010, Rov-2011} in complex oxide materials. When used in conjunction with pressure and magnetic-field tuning, Raman scattering can provide pressure- and magnetic-field-dependent information about the energy and lifetime of phonons, magnons, and spin-phonon coupling effects. In this paper, we report an inelastic light (Raman) scattering study of magnon and phonon excitations in \coo\,as simultaneous functions of temperature, pressure, and magnetic field. Below $T_C=94$ K, we report the development in \coo\,of  a\, $\sim$ $16 \,\text{cm}^{-1}$ (2 meV) $\boldsymbol q=0$ magnon excitation with \tg\,symmetry. The anomalously large Raman scattering susceptibility associated with the \tg\,symmetry magnon in \coo\,is indicative of a large magneto-optical response arising from large magnetic fluctuations that couple strongly to the dielectric response; this coupling is likely associated with the dielectric anomalies \cite{Sin-2011} observed in the ferrimagnetic phase of \coo. We also show that the Raman intensity of the \tg-symmetry magnon in \coo\,exhibits a strong suppression with increasing magnetic field, suggesting that the dramatic magneto-dielectric behavior \cite{Muf-2010, Yan-2012} observed in \coo\,results from the magnetic-field-induced suppression of magnetic fluctuations that are strongly coupled to phonons. \cite{Law-2003} Using applied pressure to increase the magnetic anisotropy in \coo\,results in a decreased magnetic field-dependence of the \tg-symmetry magnon Raman intensity in \coo, suggesting that pressure or epitaxial strain can be used to control magnetodielectric behavior and the magneto-optical response in \coo\,by suppressing magnetic fluctuations. 

\section{Experimental Methods}

\subsection{Crystal Growth and Characterization}
\coo\,crystals were grown by chemical vapor transport (CVT) following a procedure described by Ohgushi et al.\cite{Ohg-2008} Polycrystalline powder samples of \coo\,were first synthesized using cobalt nitrate hexahydrate (Strem Chemicals 99\%) and chromium nitrate nonahydrate (Acros 99\%). The nitrates were combined in stoichiometric amounts and dissolved in water. The solution was heated to $350^\circ$ C and stirred using a magnetic stir bar at $300$ rpm until all of the liquid evaporated. The remaining powder was heated in an alumina crucible at $900^\circ$ C for $16$ hours and then air quenched. Crystal samples of \coo\,were grown by CVT using CrCl$_3$ as a transport agent. $2.0$ g of polycrystalline samples and $0.04$ g of CrCl$_3$ were sealed in an evacuated quartz ampoule, which was placed inside a three-zone furnace having $950 ^\circ $C at the center with a temperature gradient of $10 ^\circ $C/cm for one month. Crystals with typical dimensions of $2 \times 2 \times 2$\,mm$^3$ were obtained.

\begin{figure}
\includegraphics[height=7.4cm,width=8.5cm]{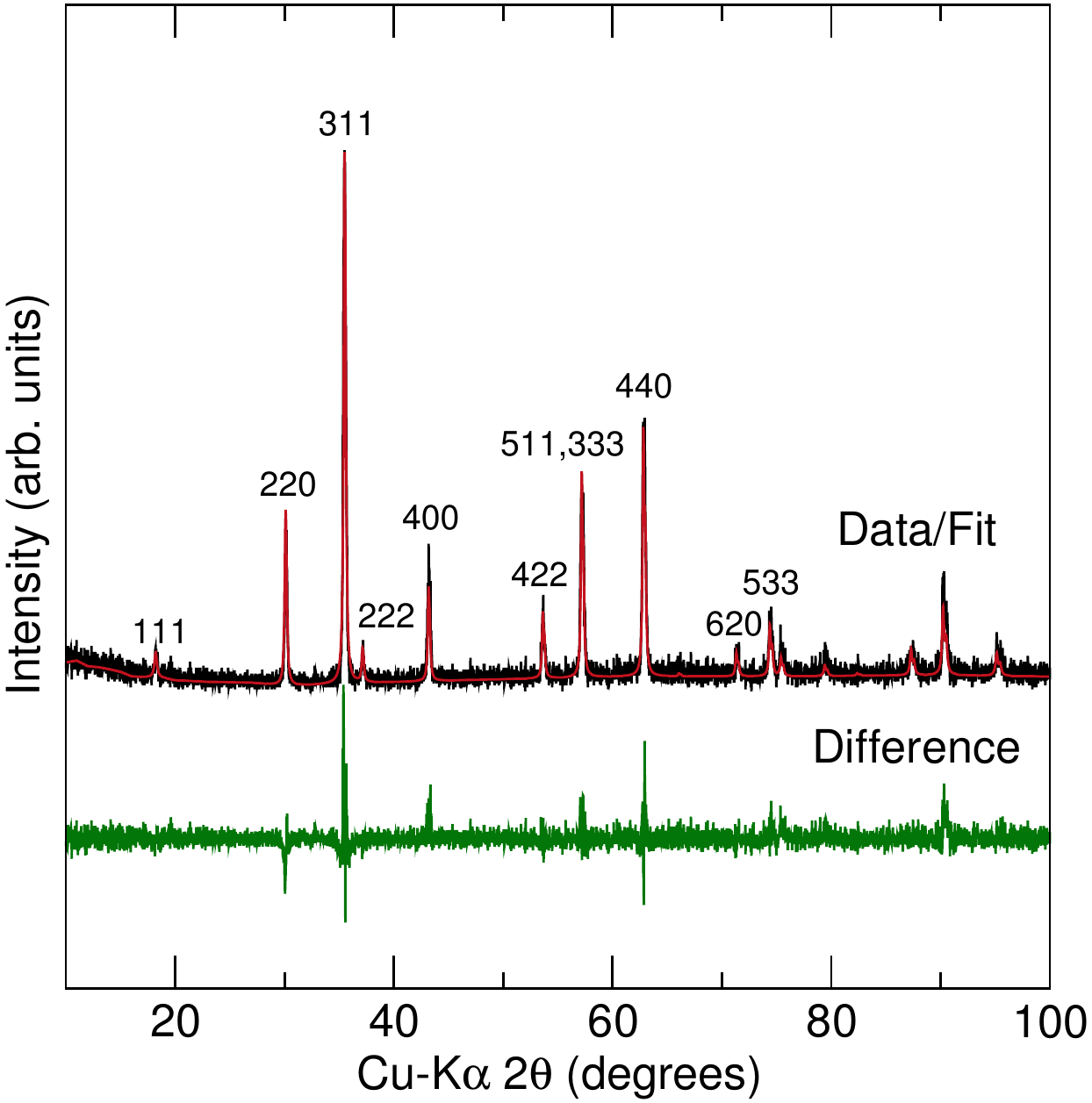}
\caption{\label{fig1}X-ray diffraction pattern and Rietveld fit of \coo\,at 298 K. The Miller indices for a cubic unit cell with cell parameter $a=8.334(1) \text{\AA}$ are also shown.}
\end{figure}

\begin{figure}
\includegraphics[scale=0.4]{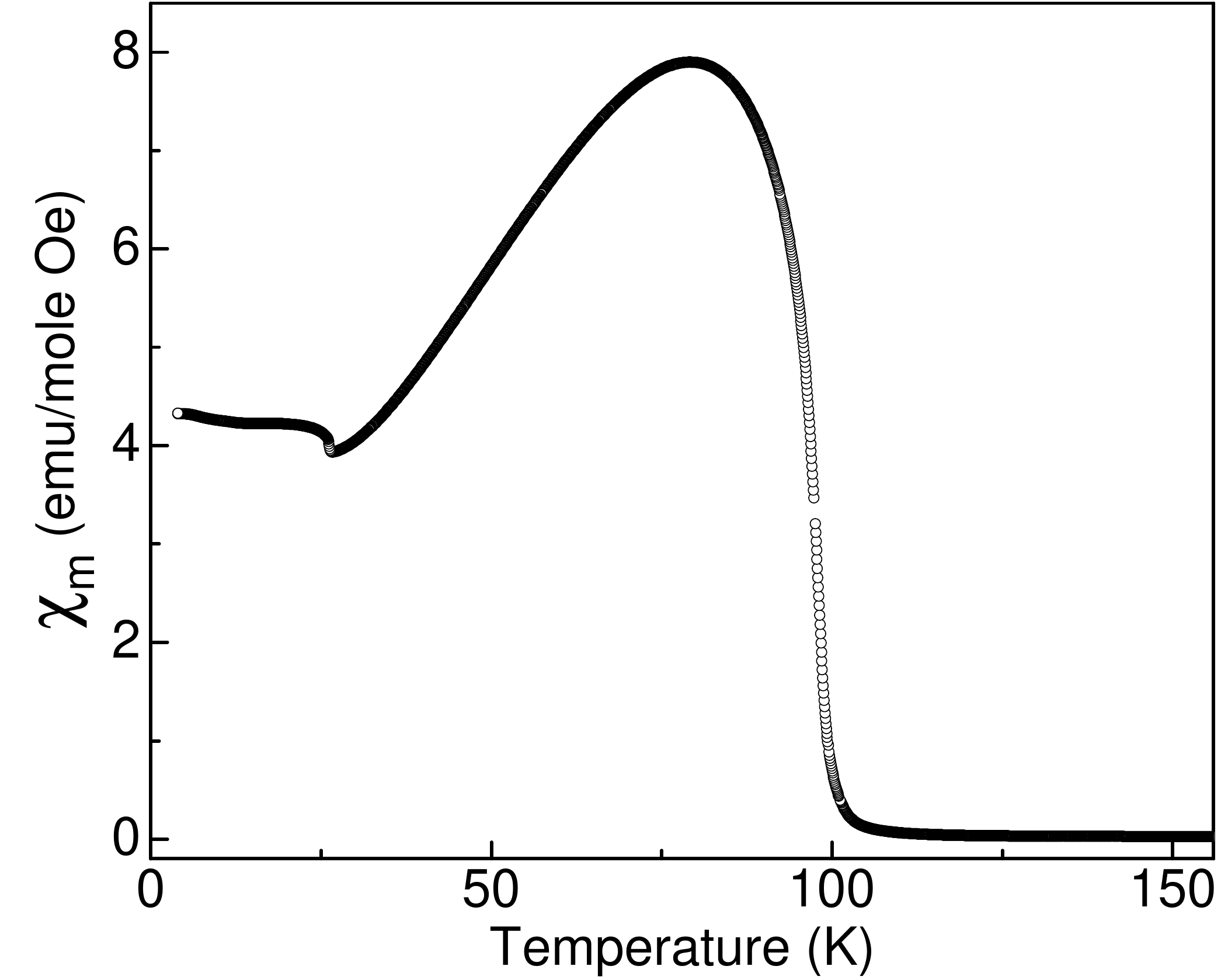}
\caption{\label{fig2}Molar susceptibility of \coo\,as a function of temperature measured in an applied field of 100 Oe.}
\end{figure}

The \coo\,crystals were characterised using x-ray diffraction and magnetization measurements. Crystals of \coo\,were ground to a powder to obtain the x-ray diffraction pattern using a Siemens-Bruker D5000 diffractometer using Cu-K$\alpha$ radiation shown in Fig.~\ref{fig1}. Rietveld refinement of the \coo\,cell to the XRD data was performed using XND Rietveld, \cite{xnd} and indicates a pure sample with $Fd \bar{3} m$ symmetry and a lattice constant of $8.334(1)$\AA , which agrees with the established structure. \cite{Tor-2012} The $<110>$ reflections from a single crystal of \coo\,were measured, and no evidence of twinning imperfections was found. The field-cooled dc magnetization data on the \coo\,powder from which our crystal sample was obtained was collected using a Quantum Design MPMS-3 and is shown as a function of temperature in Fig.~\ref{fig2}. Our results are similar to existing data. \cite{Law-2006} In particular, the sudden increase in the molar susceptibility, $\chi_{\text m}$ at $T\sim 94$ K marks the onset of ferrimagnetic ordering. The change in slope of the graph at $T\sim26$ K and an additional small anomaly at $T\sim14$ K correspond to the incommensurate and commensurate spiral ordering, respectively, in \coo. 

\subsection{Raman Scattering Measurements}
Raman scattering measurements were performed using the $647.1\,\text{nm}$ excitation line from a Kr$^+$ laser. The incident laser power was limited to $5-10$ mW, and was focused to a $\sim \text{\SI{50}{\micro\meter}}$-diameter spot to minimize laser heating of the sample. Sample heating by the laser was estimated to be in the range $5-7$ K, and this estimated laser heating is included in the temperatures given in the results section. The scattered light from the samples was collected in a backscattering geometry, dispersed through a triple stage spectrometer, and then detected with a liquid-nitrogen-cooled CCD detector. The samples were inserted into a continuous He-flow cryostat, which was horizontally mounted in the open bore of a superconducting magnet. \cite{Kim-2011} This experimental arrangement allows Raman scattering measurements under the simultaneous conditions of low temperature ($3-300$ K), high magnetic fields ($0-9$ T), and high pressures ($0-100$ kbar). To determine the symmetries of the measured Raman excitations in zero magnetic field, linearly polarized incident and scattered light were used for various crystallographic orientations of the sample. In the magnetic field measurements, circularly polarized light was used to avoid Faraday rotation of the light polarization.


\begin{figure}
\subfloat{\includegraphics[height=5.2cm,width=8.5cm]{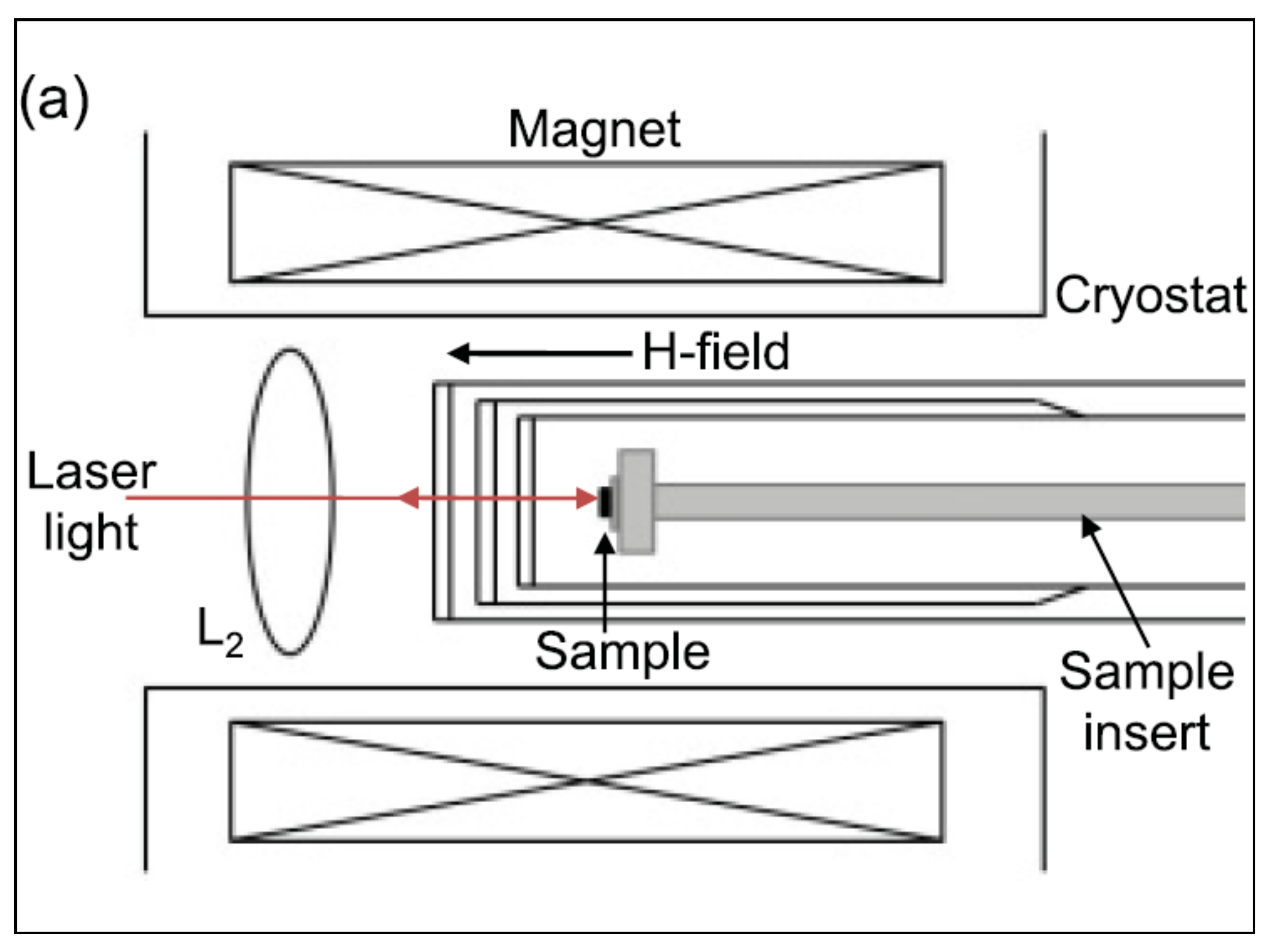} \label{exa}}\\
\subfloat{\includegraphics[height=3.8cm,width=4.1cm]{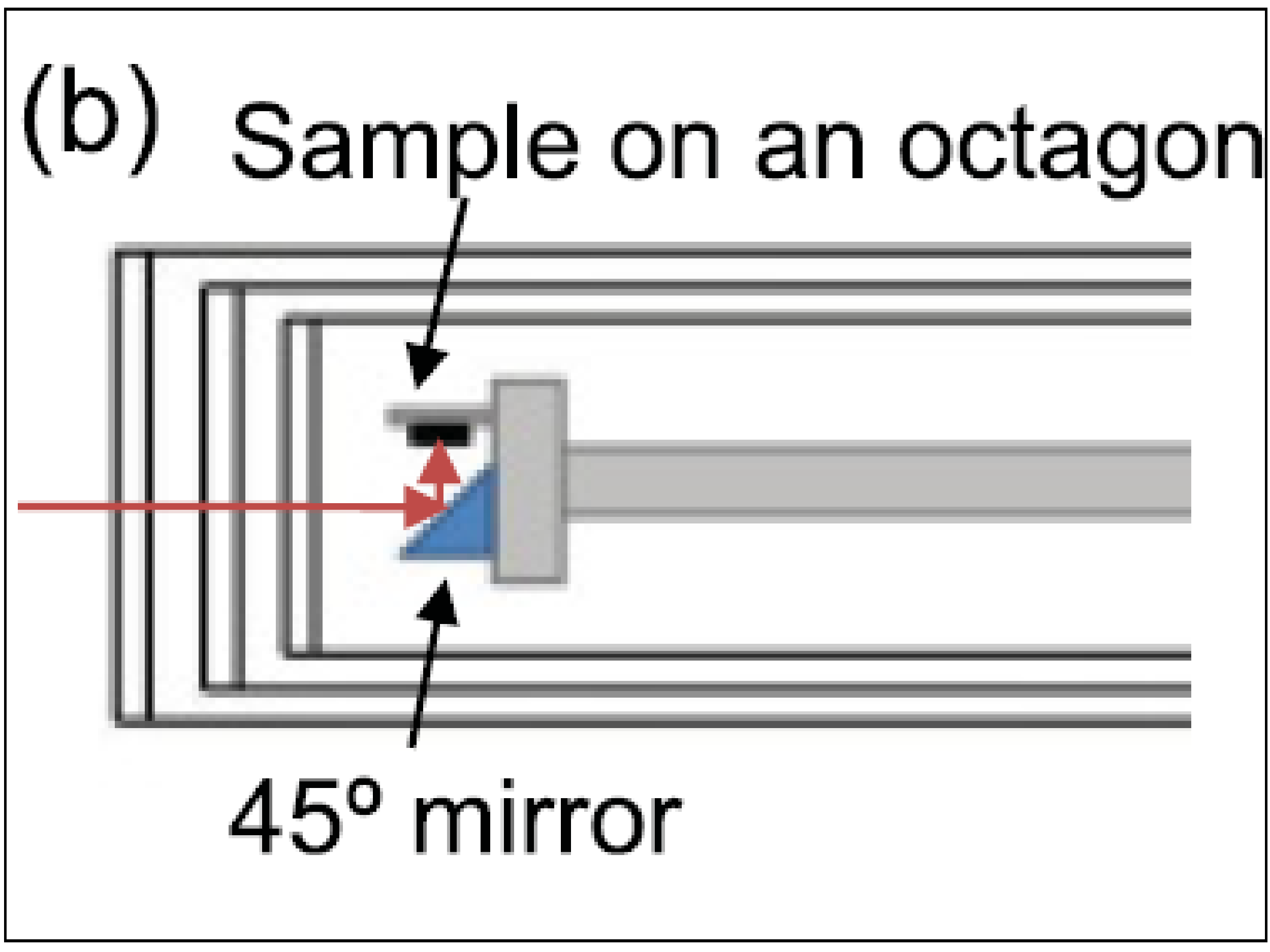} \label{exb}} 
\subfloat{\includegraphics[height=3.8cm,width=4.1cm]{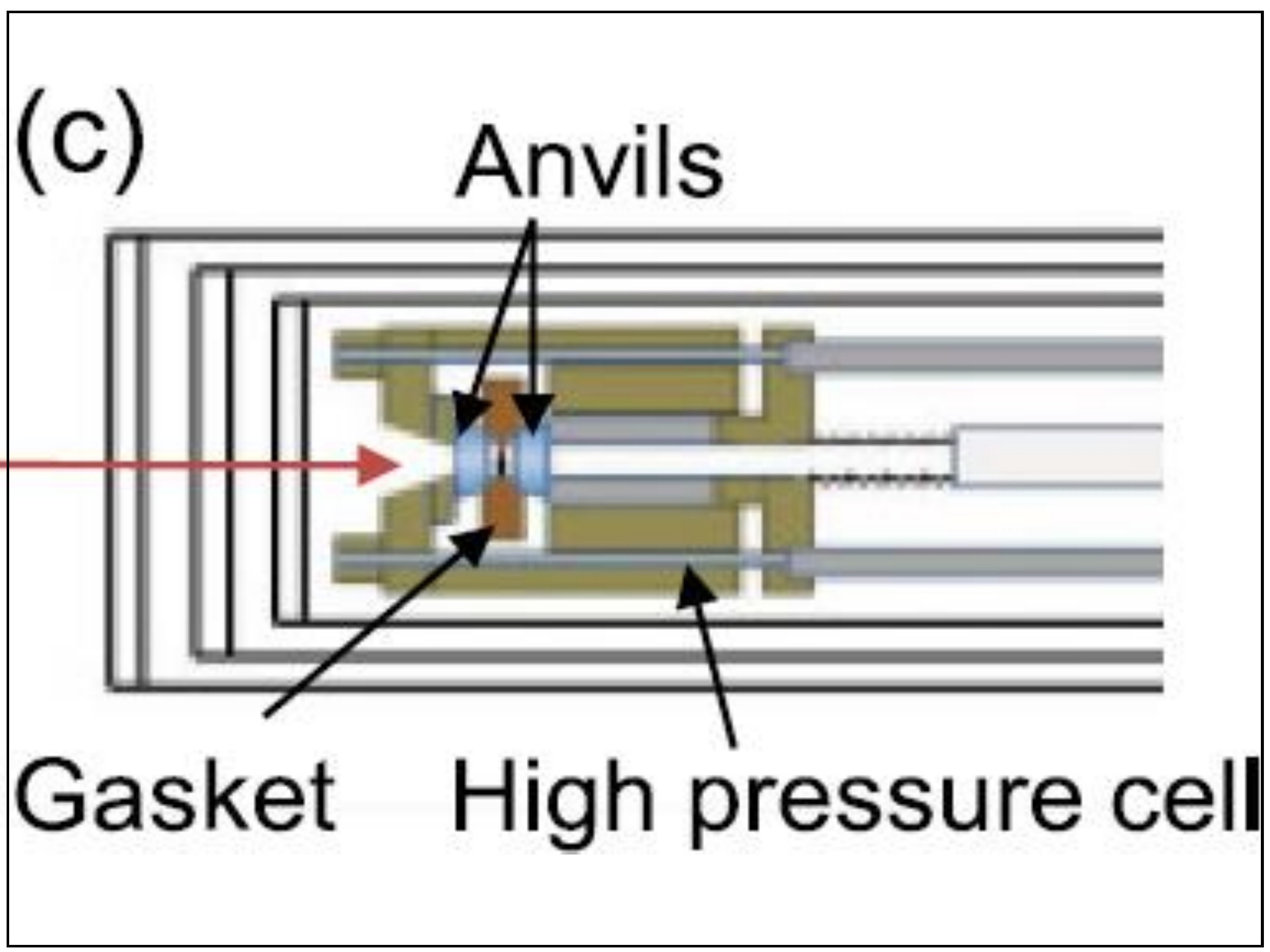} \label{exc}}
\caption{\label{fig3}Illustrations showing the experimental arrangements used for different high-magnetic-field and high temperature Raman scattering experiments at low temperatures in this study. \cite{Kim-2011} (a) Configuration for high-magnetic field measurements in the Faraday ($\boldsymbol k \parallel \boldsymbol H$) geometry, where $\boldsymbol k$ is the wavevector of the incident light and $\boldsymbol H$ is the applied magnetic field direction. (b) Configuration for high-magnetic-field measurements in the Voigt ($\boldsymbol k \perp \boldsymbol H$) geometry. (c) Configuration for high pressure measurements using a diamond anvil cell.}
\end{figure}

Magnetic field measurements were performed in both Voigt ($\boldsymbol k \perp \boldsymbol M \parallel \boldsymbol H$) and Faraday ($\boldsymbol k \parallel \boldsymbol M \parallel \boldsymbol H$) geometries, where $\boldsymbol k$ is the wavevector of the incident light and $\boldsymbol M$ is the magnetization direction. \cite{Kim-2011} Because of the very small anisotropy field in \coo\,($H_A\leq0.1$ T), \cite{Kam-2013} the net magnetization $\boldsymbol M$ was assumed to follow the applied field $\boldsymbol H$ in all experiments performed. To verify this, we confirmed that the field-dependence of the Raman spectrum was independent of the crystallographic orientation of the applied field. The field measurements in the Faraday geometry were performed by mounting the sample at the end of the insert, as illustrated in Fig.~\subref*{exa}, so that the wavevector of the incident light is parallel to the applied field. The Voigt geometry was achieved by mounting the sample on an octagon plate, which was mounted sideways on the sample rod, as illustrated in Fig.~\subref*{exb}. The incident light was guided to the sample surface with a 45$^\circ$ mirror mounted on the sample rod. This sample mounting arrangement allows the magnetic field to be applied perpendicular to the wavevector of the incident light,  $\boldsymbol k \perp \boldsymbol M \parallel \boldsymbol H$. 

 High pressure measurements were performed using a miniature cryogenic diamond anvil cell (MCDAC) to exert pressure on the sample via an argon liquid medium. The high-pressure cell was inserted into the cryostat as illustrated in Fig.~\subref*{exc}, allowing the pressure to be changed \textit{in situ} at low temperatures without any extra warming/cooling procedure. This arrangement also allows simultaneous high-pressure and high-magnetic field measurements in the Faraday  ($\boldsymbol k \parallel \boldsymbol M \parallel \boldsymbol H$) geometry, as illustrated in Fig.~\subref*{exc}. \cite{Kim-2011} The pressure was determined from the shift in the fluorescence line of a ruby chip loaded in the cell along with the sample piece.

\section{Temperature dependence of the magnetic excitation at \textit{P}=0 and \textit{B}=0}
\subsection{Results}
\begin{figure}
\includegraphics[scale=0.4]{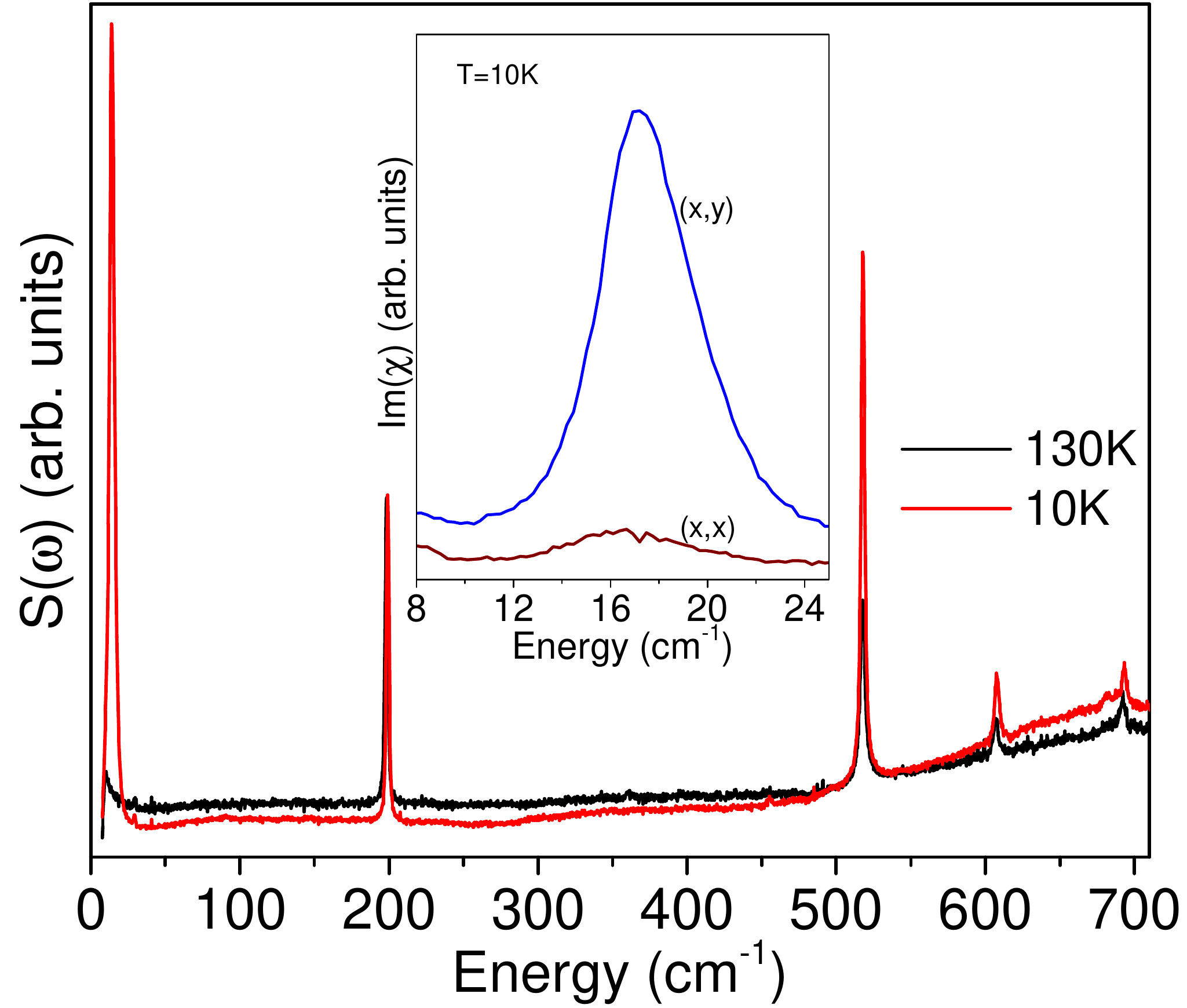}
\caption{\label{fig4}Temperature-dependence of the Raman scattering intensity, $S(\omega)$, for \coo\,at 10 K and 130 K, showing the phonon modes above 150 cm$^{-1}$ and the \tg\,symmetry magnon near 16 cm$^{-1}$ that evolves for $T<90$ K. Inset shows the polarization dependence of the magnon in \coo; the presence of this mode only in the depolarized geometry for all crystallographic orientations is indicative of the \tg\,symmetry, which transforms like an axial vector.}
\end{figure}

Fig.~\ref{fig4} shows the $T=10$ K and $T=130$ K Raman spectra of \coo\,between 0-700 cm$^{-1}$ in a scattering geometry with circularly polarized incident light and unanalyzed scattered light. The $T=10$ K spectrum exhibits the five Raman-active phonon modes expected and previously observed \cite{Kus-2009, Pta-2014, Eft-2015} for \coo, including phonon modes at 199 cm$^{-1}$, 454 cm$^{-1}$, 518 cm$^{-1}$, 609 cm$^{-1}$, and 692 cm$^{-1}$ (at $T=10$ K). In addition to the phonon modes, the $T=10$ K spectrum in Fig.~\ref{fig4} has an additional mode that develops near 16 cm$^{-1}$ ($\sim$2 meV) below $T=90$ K. The inset of Fig.~\ref{fig4} shows that the 16 cm$^{-1}$ mode intensity is present only in the ``depolarized" scattering geometry, \textit{i.e.}, only when the incident and scattered light polarizations are perpendicular to one another, independent of the crystallographic orientation. This polarization dependence indicates that the 16 cm$^{-1}$ mode symmetry transforms like the fully antisymmetric representation, \tg, which has the symmetry properties of an axial vector, characteristic of a magnetic excitation. \cite{Cot-1986, Ram-1991} Consequently, we identify the 16 cm$^{-1}$ excitation as a $\boldsymbol q=0$ \tg \,symmetry magnon in \coo. This interpretation is supported by the temperature-dependence of the 16 cm$^{-1}$ \tg -symmetry mode Raman scattering susceptibility, $\text{Im}\,\chi(\omega)$ (see Fig.~\subref*{T2a}), where $\text{Im}\,\chi(\omega)=S(\boldsymbol q=0,\omega)/[1+n(\omega,T)]$, $S(\boldsymbol q=0,\omega)$ is the measured Raman scattering response, and $[1+n(\omega,T)]$ is the Bose thermal factor with $n(\omega,T)=[e^{\hbar\omega/k_BT})-1]^{-1}$. Fig.~\subref*{T2b} shows that the $\sim16\,\text{cm}^{-1}$ \tg\,symmetry mode energy (solid squares) decreases in energy (``softens") with increasing temperature toward $T_C$\textemdash consistent with the temperature-dependence of the Co$^{2+}$ sublattice magnetization \cite{Kam-2013}\textemdash indicative of a single-magnon excitation. \cite{Cot-1986} Fig.~\subref*{T2b} also shows that the amplitude of the Raman susceptibility, $\text{Im}\,\chi(\omega)$, associated with the 16 cm$^{-1}$ \tg -symmetry magnon mode (solid circles) is relatively insensitive to temperature and is  comparable to that of the 199 cm$^{-1}$ T$_{2g}$ phonon. Notably, the 16 cm$^{-1}$ \tg\,symmetry magnon we observe in \coo\,has a similar energy and temperature dependence to that of the exchange magnon observed previously in terahertz \cite{Tor-2012} and infrared spectroscopy \cite{Kam-2013} measurements of \coo. Nevertheless, it is unlikely that the 16 cm$^{-1}$ \tg\, symmetry magnon we observe in \coo\,is the same as the intersublattice exchange mode reported in infrared measurements, because \tg\,is not an infrared-active symmetry. Note in this regard that the spinel structure of \coo\,is expected to exhibit six $\boldsymbol q=0$ magnon modes with 5 closely spaced optical branches, \cite{Kap-1953, Bri-1966, Sah-1974, Tor-2012} so we are likely observing a different optical magnon that is close in energy to that observed in infrared measurements. \cite{Tor-2012, Kam-2013}

\subsection{Discussion and Analysis}
\begin{figure}
\subfloat{\includegraphics[height=3.6cm,width=3.6cm]{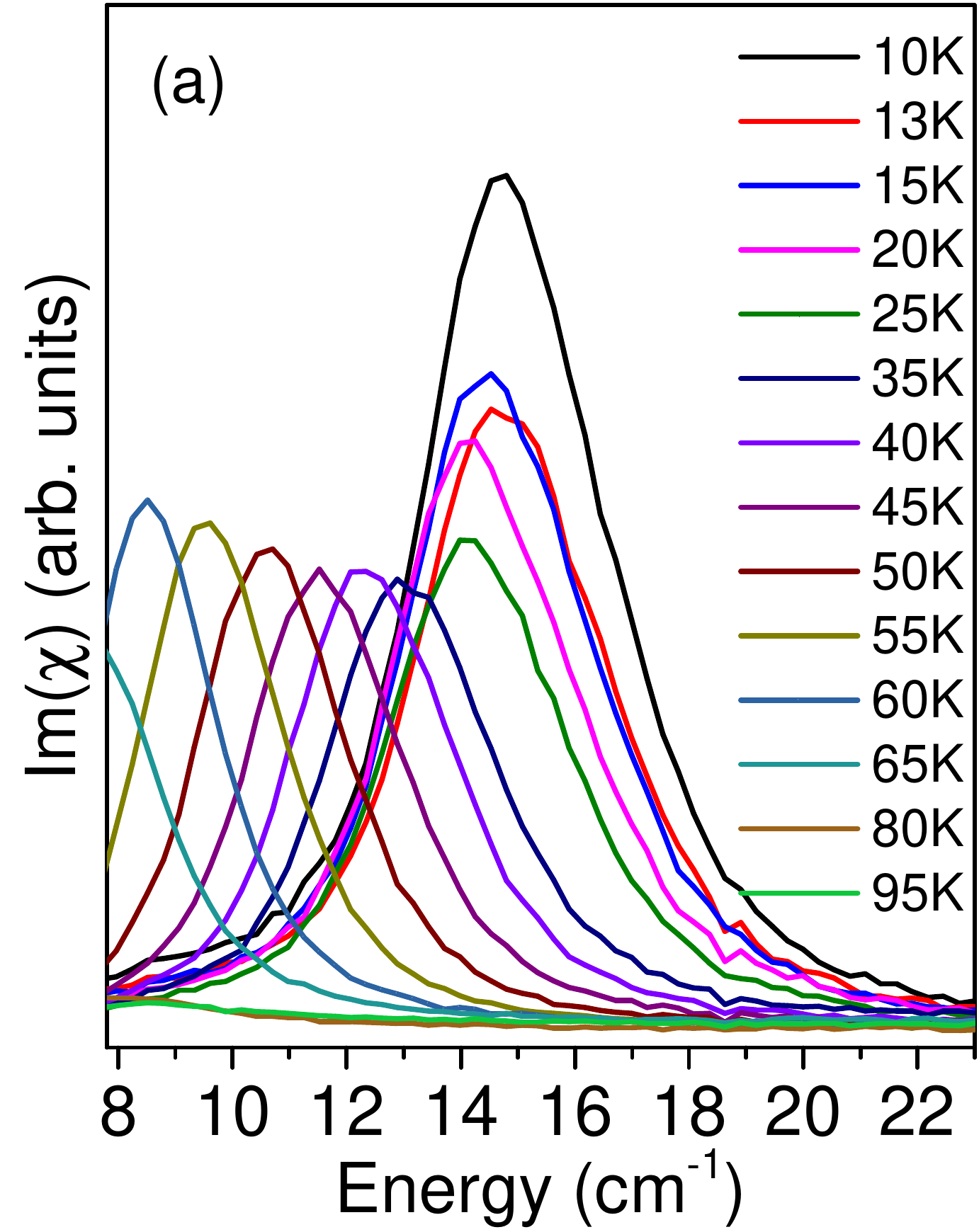}
\label{T2a}}
\quad
\subfloat{\includegraphics[height=3.6cm,width=4cm]{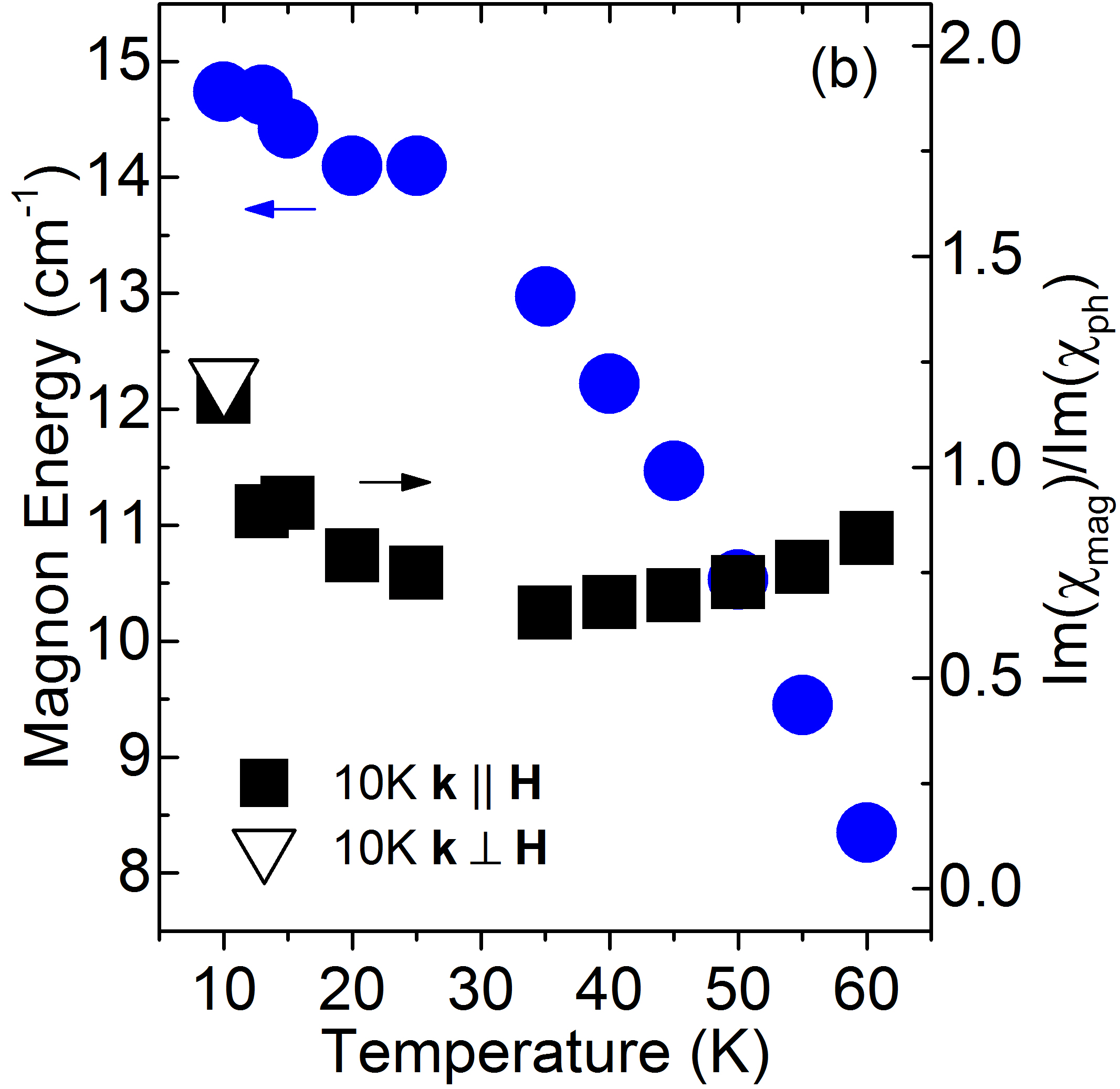}
\label{T2b}}
\caption{\label{fig5}(a) Raman scattering susceptibility, $\text{Im}\,\chi(\omega)$, of the \tg-symmetry magnon of \coo\,as a function of temperature. (b) Summary of the temperature dependence of the \tg\,symmetry magnon energy (filled squares). Also shown in filled circles is a summary of the temperature dependence of the \tg\,symmetry magnon Raman susceptibility amplitude normalized to the susceptibility amplitude of the 199 cm$^{-1}$ T$_{2g}$ optical phonon, $\text{Im}\,\chi_{mag}(\omega)/\text{Im}\,\chi_{ph}(\omega)$.}
\end{figure}

The finite $\boldsymbol q=0$ energy of the $\omega \sim$ 16 \text{cm}$^{-1}$ (2 meV) \tg-symmetry magnon in \coo\,primarily reflects the finite exchange, $H_E$, and anisotropy, $H_A$, fields in \coo, according to $\omega=\gamma(2H_AH_E + {H_A}^2)^{1/2}$, where $\gamma$ is the gyromagnetic ratio $g\mu_B/\hbar$. \cite{Ram-1991} Fig. ~\ref{fig5} also shows that the 16 cm$^{-1}$ \tg\, symmetry magnon in \coo\,is apparent to temperatures as high as $T\sim60$ K, indicating that the \tg\,symmetry magnon in \coo\,is dominated by the Co$^{2+}$ sublattice spins, which order at a significantly higher temperature (94 K) than the Cr$^{3+}$ sublattice (49 K). \cite{Kam-2013}

Importantly, the Raman susceptibility of the 16 cm$^{-1}$ \tg\, symmetry magnon at T=10 K (for $H=0$ T and $P=0$ kbar) (see Fig.~\ref{fig5}) reflects the degree to which this magnon modulates the dielectric response, $\epsilon=4\pi\chi_E$ (where $\chi_E$ is the electric susceptibility). \cite{Dem-1987, Kum-2011} Consequently, while Raman scattering from magnons is generally much weaker than Raman scattering from phonons, \cite{Cot-1986} Figs.~\ref{fig4} and \ref{fig5} show that Raman intensity of the \tg-symmetry magnon is comparable to that of the Raman-active phonons in \coo, indicative of a strong influence of this magnon on the dielectric response of \coo.

The large Raman susceptibility of the \tg\, symmetry magnon reflects a large magneto-optical response in \coo, and is likely associated with strong magnetic fluctuations that modulate the dielectric response via strong spin-lattice coupling. \cite{Bar-1983, Dem-1987, Bor-1988} Such large magnetic fluctuations are attributable to the weak anisotropy field in \coo, $H_A\leq$0.1 T, \cite{Kam-2013} and can contribute in several ways to fluctuations in the dielectric response:\cite{Bar-1983, Dem-1987, Bor-1988}

\begin{equation}
\delta \epsilon(m, l) = i \, f \delta m + g (\delta l )^2 + a (\delta m)^2 
\label{eq1}
\end{equation}
where $\delta\epsilon$ is the dielectric response fluctuation, $\delta m$ represents longitudinal fluctuations in the magnetization, $\delta l$ represents fluctuations of the antiferromagnetic vector, and $a$, $f$, and $g$ are constants. The first term in Eq.~\eqref{eq1} is associated with the linear magneto-optical Faraday effect, the second term is associated with linear magnetic birefringence, and the final term is an isotropic ``exchange" mechanism for magnon scattering that is present in non-collinear antiferromagnets. \cite{Dem-1987, Vit-1991} In non-collinear antiferromagnetic and ferrimagnetic materials with weak anisotropy\textemdash such as \coo\textemdash strong single-magnon scattering can result from large fluctuations of both $l$ and $m$. In particular, the one-magnon Raman scattering intensity, $S$, associated with large magnetic fluctuations of the antiferromagnetic vector at $H=0$ is limited only by the anisotropy field, $H_A$ (i.e., $S \propto 1/H_A$), \cite{Dem-1987} which is very small in \coo, $H_A\leq$0.1 T. \cite{Kam-2013}

\section{Magnetic-field-dependence of the \tg-symmetry magnon in \coo}
\subsection{Results}
\begin{figure}
\subfloat{\includegraphics[scale=0.25]{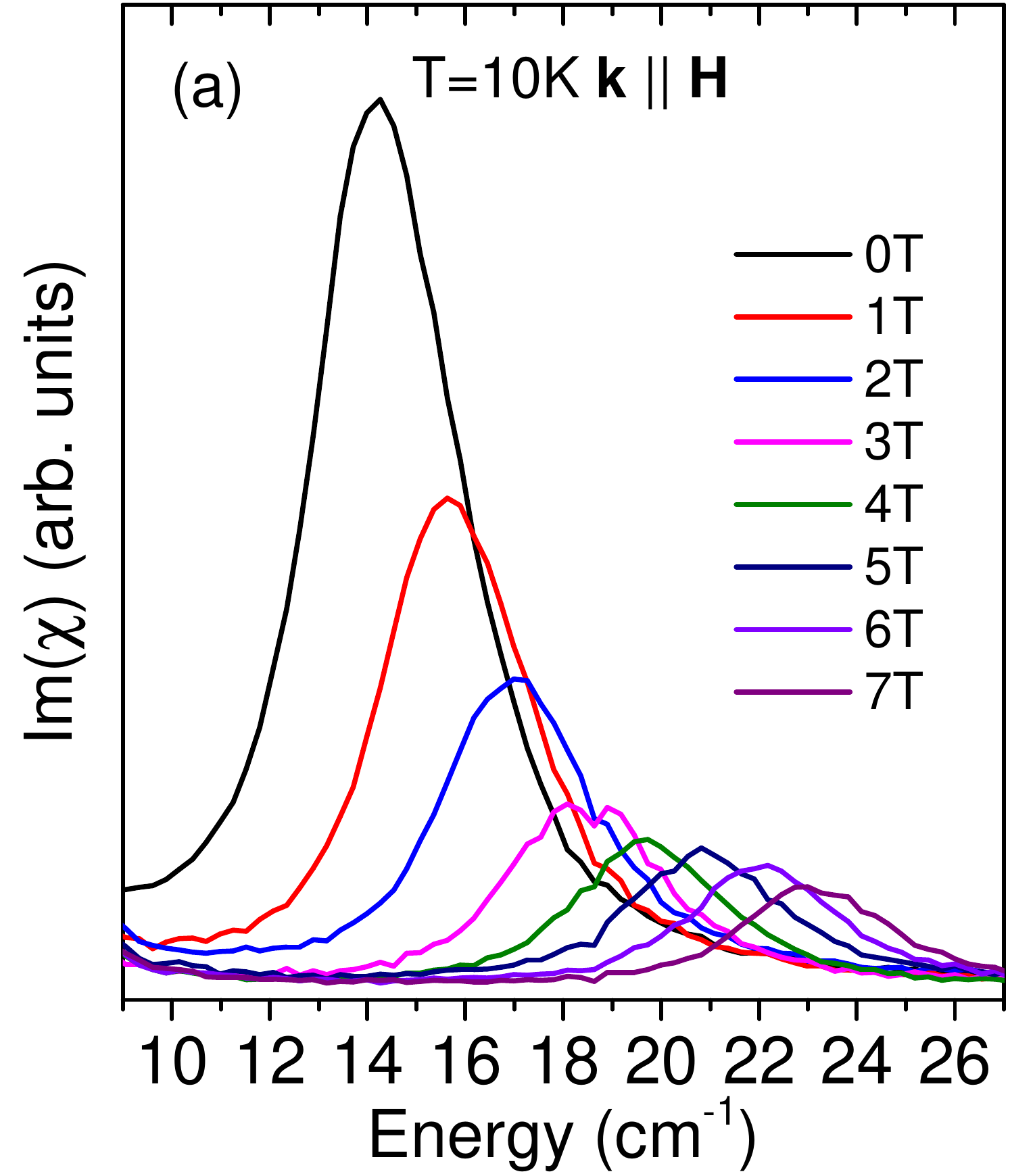}
\label{F1a}}
\quad
\subfloat{\includegraphics[scale=0.25]{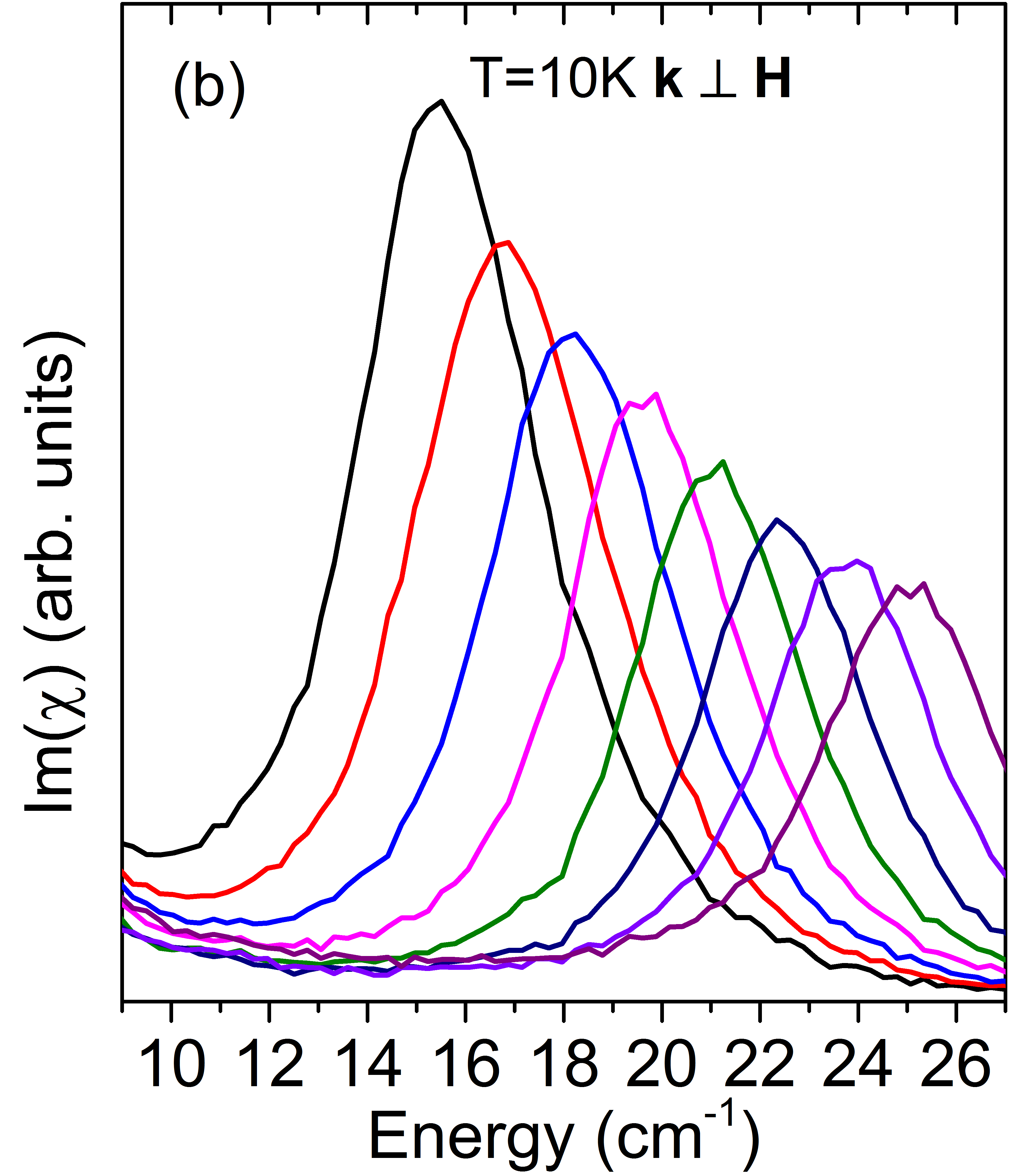}
\label{F1b}}
\\
\subfloat{\includegraphics[scale=0.24]{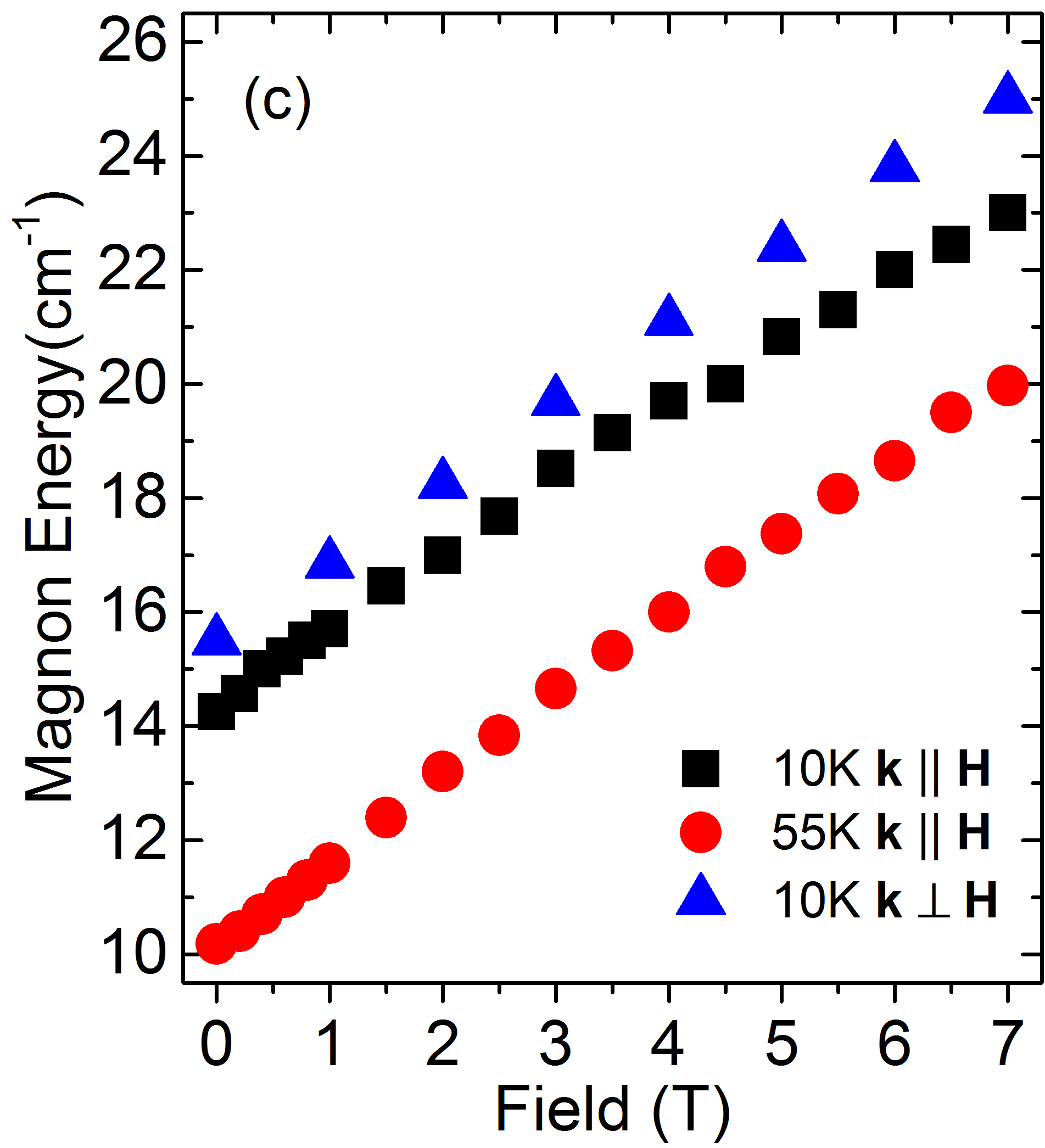}
\label{F1c}}
\quad
\subfloat{\includegraphics[scale=0.24]{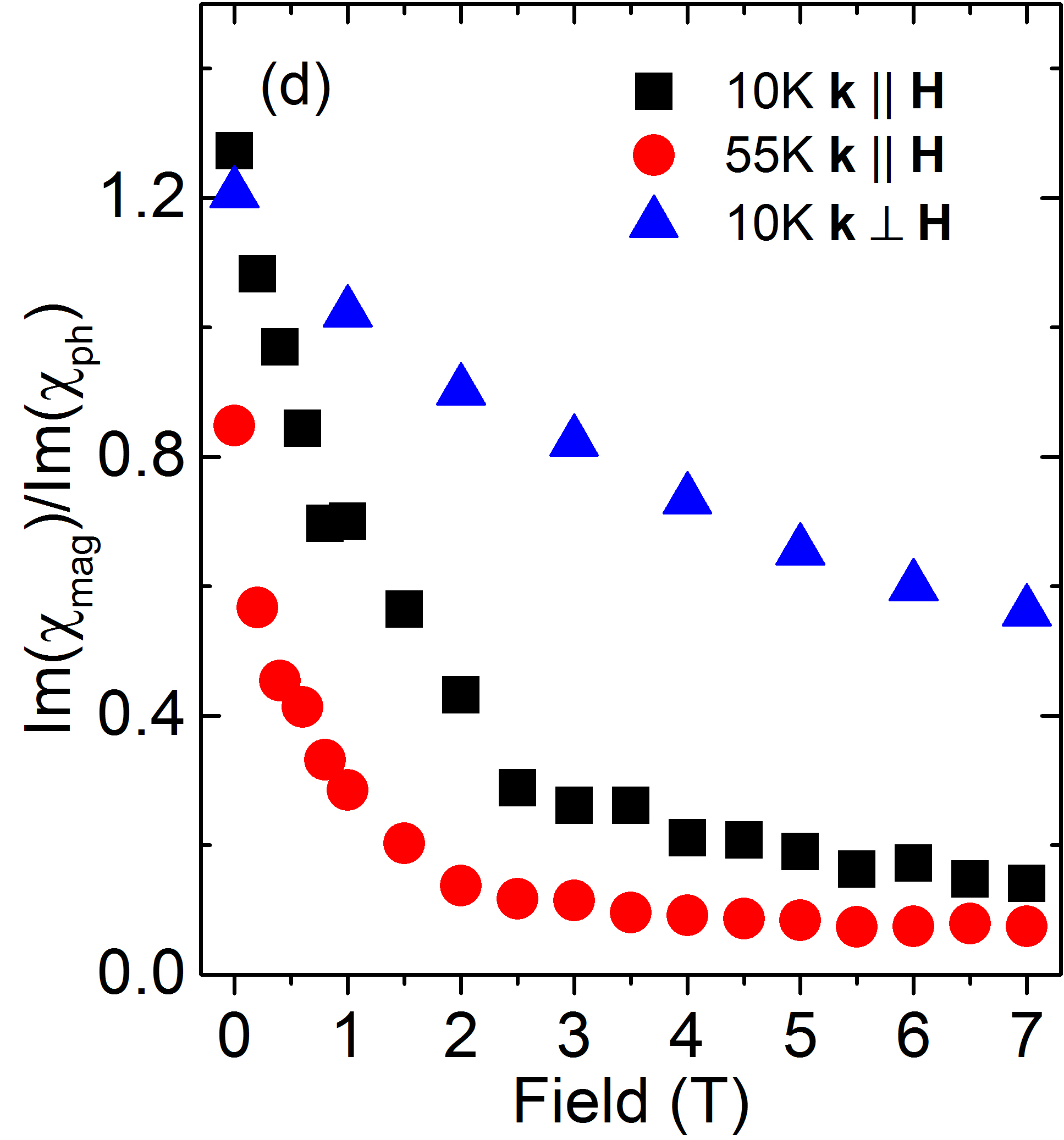}
\label{F1d}}
\caption{\label{fig6} Magnetic-field-dependence of the Raman scattering susceptibility, $\text{Im}\,\chi(\omega)$, of the \tg-symmetry magnon in \coo\,at $T=10$ K in the (a) Faraday geometry ($\boldsymbol k\parallel\boldsymbol M\parallel\boldsymbol H$) and the (b) Voigt geometry ($\boldsymbol k \perp \boldsymbol M \parallel \boldsymbol H$). (c) Summary of the field dependences of the \tg-symmetry magnon energy of \coo\,at (filled squares) $T=10$ K and (filled circles) $T=55$ K in the Faraday geometry and at (filled triangles) $T=10$ K in the Voigt geometry. (d) Summary of the field dependences of the amplitude of the \tg-symmetry magnon Raman susceptibility normalized to the amplitude of the 199 cm$^{-1}$ T$_{2g}$ phonon Raman susceptibility at (filled squares) $T=10$ K and (filled circles) $T=55$ K in the Faraday geometry and at (filled triangles) $T=10$ K in the Voigt geometry.}
\end{figure}

Fig.~\ref{fig6} shows the magnetic-field-dependence of the Raman susceptibility, $\text{Im}\,\chi(\omega)$, for the \tg-symmetry magnon of \coo\,at $P=0$ kbar and $T=10$ K with an applied magnetic field in both the (Fig.~\subref*{F1a}) Faraday ($\boldsymbol k\parallel\boldsymbol M \parallel \boldsymbol H$) and (Fig.~\subref*{F1b}) Voigt ($\boldsymbol k \perp\boldsymbol M \parallel \boldsymbol H$) geometries. Fig.~\subref*{F1c} summarizes the field-dependences of the \tg-symmetry magnon energy at both $T=10$ K and $T=55$ K, showing that the \tg-symmetry magnon energy exhibits a linear increase with increasing field. The shift in the \tg-symmetry magnon energy with field, $d\omega /dH \sim$ 1.1 cm$^{-1}/$T corresponds to a dimensionless ratio $\hbar\omega/\mu_BH=2.4$. This ratio is close to the $T=4$ K value of $\hbar\omega/\mu_BH=2.5$  measured for the exchange magnon in \coo\,\cite{Kam-2013} and is consistent with the gyromagnetic ratio of 2.2 for Co$^{2+}$. \cite{Alt-1974, Tor-2012} Fig.~\subref*{F1d} compares the field-dependence of the normalized \tg-symmetry magnon intensity, $\text{Im}\,\chi_{mag}(\omega)/\text{Im}\,\chi_{ph}(\omega)$, in both the (filled circle and square) Faraday ($\boldsymbol k \parallel \boldsymbol M \parallel \boldsymbol H$) and (filled triangle) Voigt ($\boldsymbol k \perp \boldsymbol M \parallel \boldsymbol H$) geometries, where $\text{Im}\,\chi_{mag}(\omega)$ and $\text{Im}\,\chi_{ph}(\omega)$ are the Raman susceptibilities of the \tg-symmetry magnon and 199 cm$^{-1}$ T$_{2g}$ phonon, respectively. Fig.~\subref*{F1d} shows that there is a substantial decrease in the normalized \tg-symmetry magnon intensity of \coo\,with increasing field in both the Faraday ($\boldsymbol k \parallel \boldsymbol M \parallel \boldsymbol H$) and Voigt ($\boldsymbol k \perp \boldsymbol M \parallel \boldsymbol H$) geometries at $T=10$ K and $T=55$ K. Note that the field-dependent decrease we observe in the \tg-symmetry magnon intensity \textemdash which is particularly dramatic in the Faraday geometry ($\boldsymbol k \parallel \boldsymbol M \parallel \boldsymbol H$) \textemdash cannot be attributed to field-dependent changes in polarization or crystallographic orientation: the use of circularly polarized incident light in these experiments precludes field-dependent rotation of the incident polarization; and \tg-symmetry modes appear in the depolarized scattering geometry independent of the crystallographic orientation of the sample.

\subsection{Discussion and Analysis}
The anomalously large decrease in the 16 cm$^{-1}$ \tg-magnon Raman intensity with increasing field in the Faraday geometry ($\boldsymbol k \parallel \boldsymbol M \parallel \boldsymbol H$) of \coo\,(see Fig.~\ref{fig6}) is quite different than the field-independent magnon Raman intensities observed in other spinel materials, such as Mn$_3$O$_4$ and MnV$_2$O$_4$. \cite{Gle-2014} To clarify the anomalously strong field-dependence of the \tg-symmetry magnon Raman intensity in \coo, note that the magnon Raman intensity in the Faraday geometry is expected to be dominated by the linear magnetic birefringence contribution to dielectric fluctuations, $\delta\epsilon=g (\delta l )^2$. \cite{Dem-1987, Bor-1988, Vit-1991} Thus, the strong decrease in the \tg-symmetry magnon Raman intensity in the Faraday geometry likely reflects a field-induced decrease in fluctuations of the antiferromagnetic vector, $\delta l$. A similar field-dependent decrease in the single-magnon inelastic light scattering response associated with fluctuations of the antiferromagnetic vector was also observed in the canted antiferromagnet EuTe. \cite{Dem-1987}

Fig.~\subref*{F1b}, \subref*{F1d} shows that there is a similar, albeit less dramatic, field-dependent decrease in the \tg-symmetry magnon Raman intensity measured in the Voigt ($\boldsymbol k \perp \boldsymbol M \parallel \boldsymbol H$) geometry. This geometry is primarily sensitive to the Faraday ($\delta\epsilon=i \, f \delta m$) and isotropic exchange ($\delta\epsilon=a (\delta m)^2$) contributions to dielectric fluctuations, which are dominated by longitudinal fluctuations in the magnetization. \cite{Dem-1987, Bor-1988, Vit-1991} Altogether, the suppression of the \tg-symmetry magnon Raman scattering intensities in both Faraday and Voigt geometries is indicative of a field-induced suppression of both transverse and longitudinal magnetic fluctuations in \coo.

The field-dependent suppression of the \tg-symmetry magnon Raman intensity in \coo\,points to a specific microscopic mechanism for the magnetodielectric response observed in \coo. \cite{Sin-2011, Muf-2010, Yan-2012} Lawes \textit{et al.} have pointed out that the field-induced suppression of magnetic fluctuations can contribute to the magnetodielectric response of a material via the coupling of magnetic fluctuations to optical phonons. \cite{Law-2003} This spin-phonon coupling contributes to the magnetodielectric response of a material through field-induced changes to the net magnetization. \cite{Smo-1982, Kim-2003b, Law-2003, Yan-2012} A simple phenomenological description for how the magnetization of a magnetoelectric material influences the dielectric response of the material is obtained by considering the free energy, $F$, in a magnetoelectric material with a coupling between the magnetization $\boldsymbol M$ and polarization $\boldsymbol P$: \cite{Smo-1982, Kim-2003b, Yan-2012}

\begin{eqnarray}
F(M,P) = F_0 + aP^2 &+& bP^4 - PE + cM^2 + \nonumber 
\\  &&dM^4 - MH + eM^2P^2 ,
\label{eq2}
\end{eqnarray}
where $F_0$, $a$, $b$, $c$, $d$, and $e$ are temperature-dependent constants, and $M$, $P$, $E$, and $H$ are the magnitudes of the magnetization, polarization, applied electric field, and applied magnetic field, respectively. The dependence of the dielectric response on magnetization in a magnetoelectric material, $\epsilon(M)$ , can be obtained from the second derivative of the free energy with respect to polarization $P$: \cite{Smo-1982, Kim-2003b, Yan-2012}

\begin{equation}
[\epsilon(M)]^{-1} \sim (\partial^2F/\partial P^2) = 2a + 12bP^2 + 2eM^2,
\end{equation}
which, for a negligible macroscopic polarization $P$ in the material, can be written: \cite{Kim-2003b, Yan-2012}
\begin{equation}
\epsilon(M) = 1/[2a + 2eM^2].
\end{equation}
Thus, the dielectric response, $\epsilon = 4\pi\chi_E$, decreases with increasing squared magnetization, $M^2$ and decreasing magnetic fluctuations. \cite{Law-2003}

The above results suggest that both the magnetic-field-dependent decrease in the intensity of the 16 cm$^{-1}$ \tg-symmetry magnon (see Fig.~\ref{fig6}) and the magnetodielectric response, $\Delta\epsilon(H)/\epsilon(0)=[\epsilon(H)-\epsilon(0)]/\epsilon(0)$, in \coo\,\cite{Muf-2010, Yan-2012} reflect magnetic-field-induced changes to magnetic fluctuations\textemdash particularly fluctuations associated with the antiferromagnetic vector\textemdash that are strongly coupled to phonons \cite{Law-2003} via the biquadratic contribution to the free energy, $M^2P^2$ (see Eq.~\eqref{eq2}).

\section{Pressure dependence of the \tg-symmetry magnon in \coo}

\subsection{Results}

As discussed above, the strong \tg-symmetry magnon Raman intensity of \coo\,is believed to reflect strong magnetic fluctuations that are coupled to long-wavelength phonons, which should also be associated with significant magneto-optical responses (both linear Faraday and linear magnetic birefringence) in \coo. Our results show that the application of a magnetic field suppresses these fluctuations, leading to the substantial magnetodielectric response observed in \coo. An alternative approach to suppressing magnetic fluctuations is to use applied pressure or strain to increase the crystalline anisotropy of \coo. To investigate this possibility, magnetic-field-dependent measurements of the \tg-symmetry magnon in \coo\,were performed for different applied pressures.

\begin{figure}
\subfloat{\includegraphics[scale=0.23]{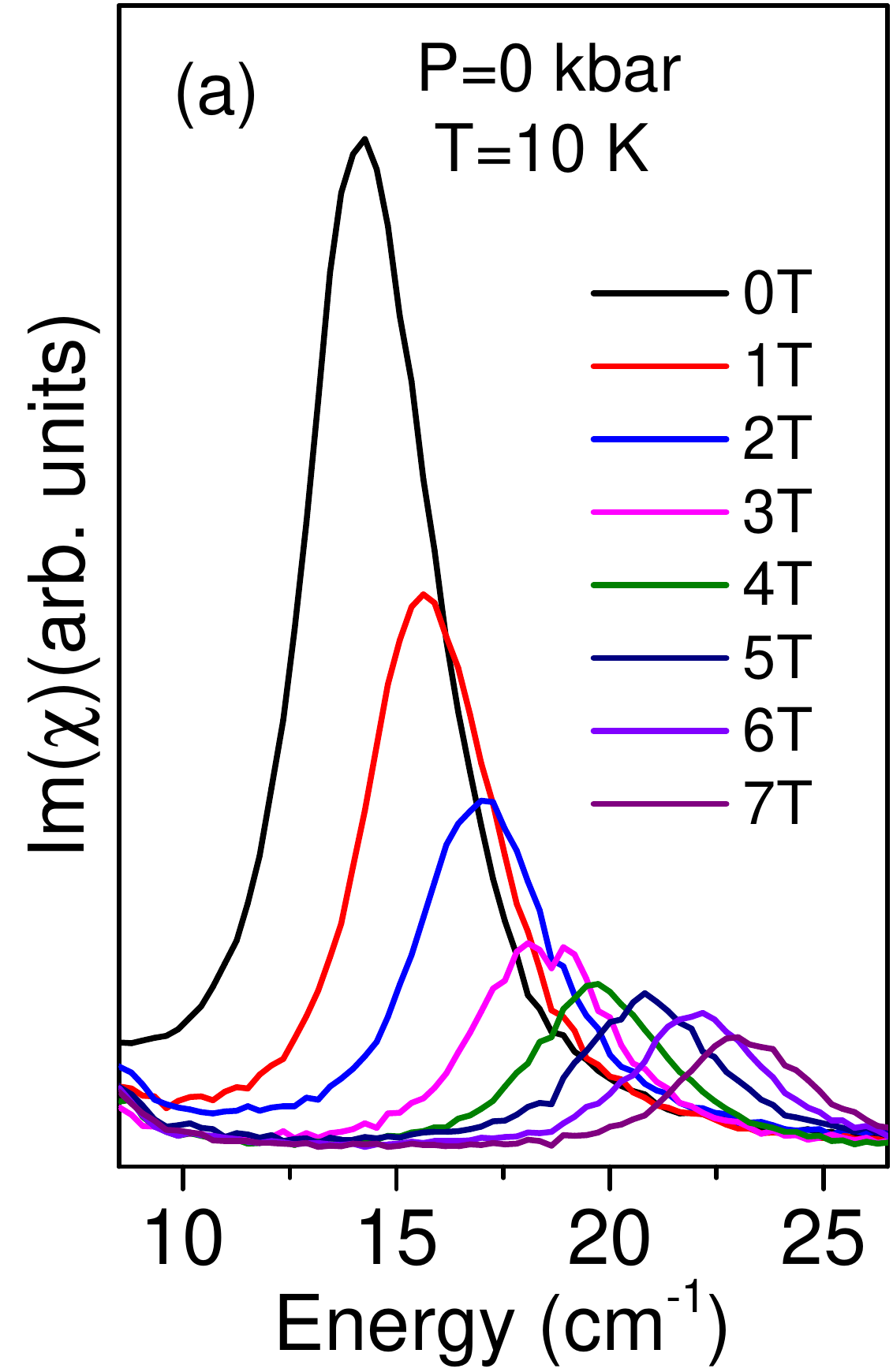}
\label{P1a}}
\,\,
\subfloat{\includegraphics[scale=0.23]{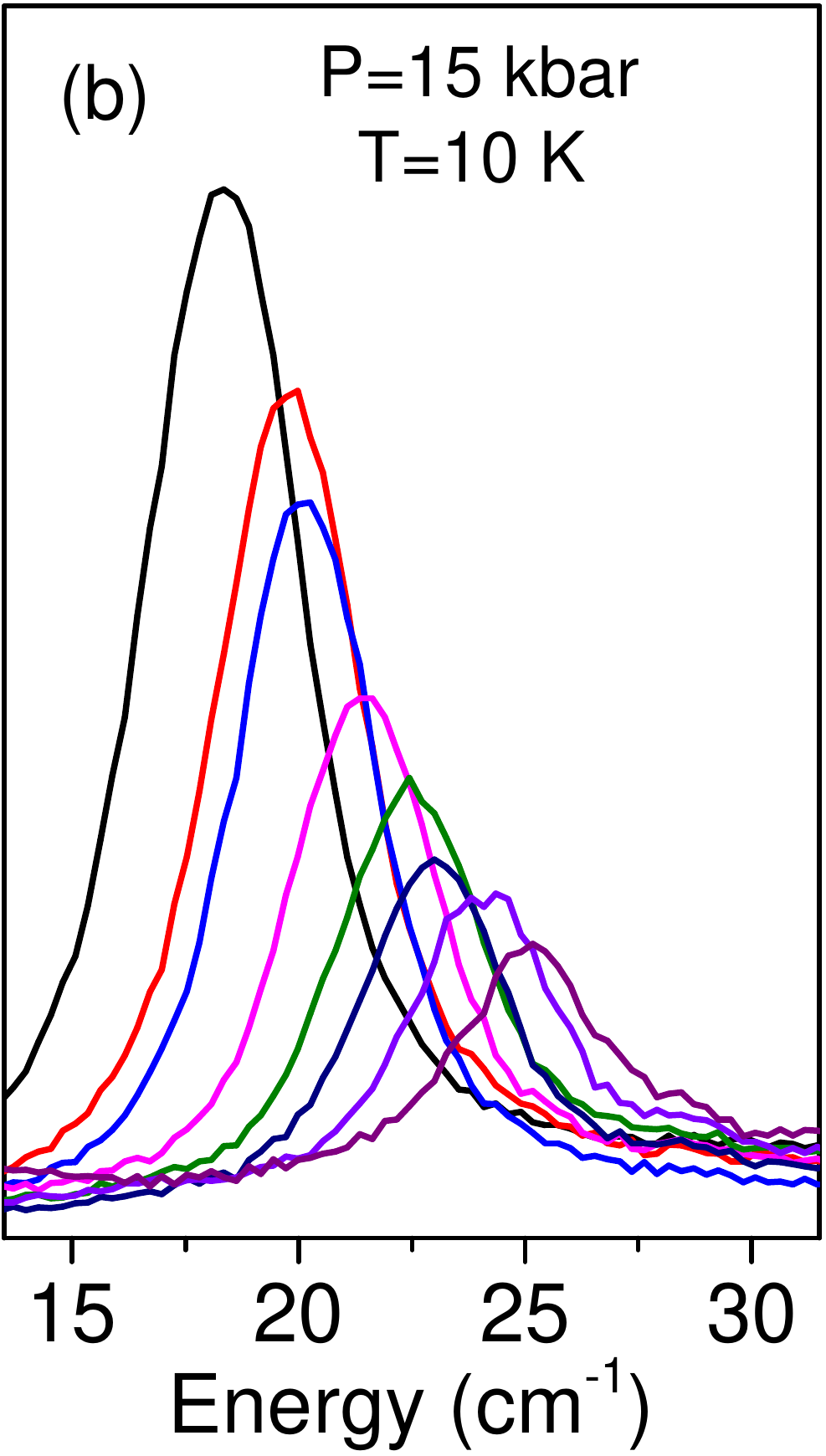}
\label{P1b}}
\,\,
\subfloat{\includegraphics[scale=0.23]{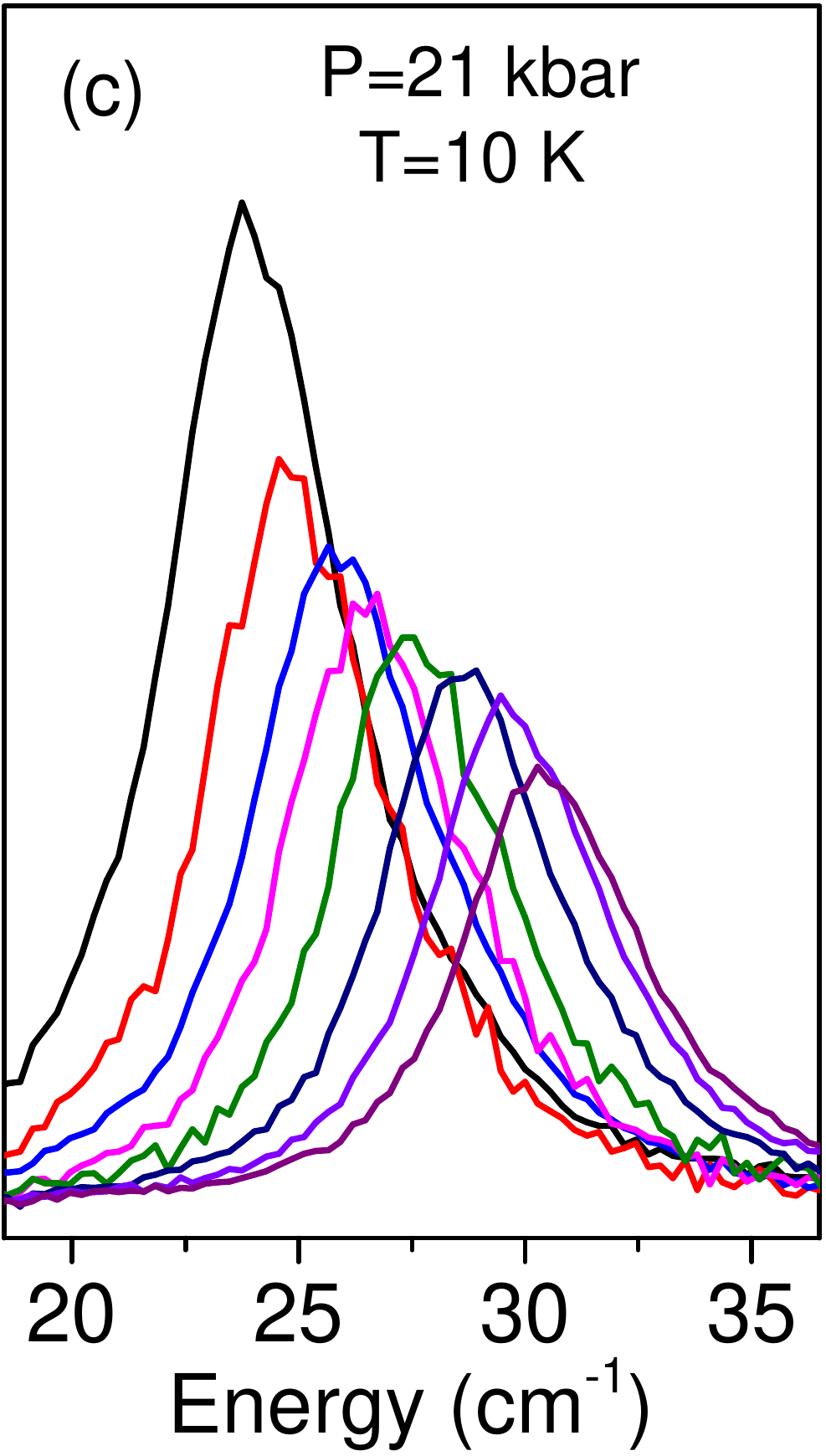}
\label{P1c}}
\caption{\label{fig7}Field-dependence in the Faraday ($\boldsymbol k \parallel \boldsymbol M \parallel \boldsymbol H$) geometry of the \tg-symmetry magnon Raman susceptibility of \coo\,at $T=10$ K and at various applied pressures, including (a) $P=0$ kbar, (b) $P=15$ kbar, and (c) $P=21$ kbar.}
\end{figure}

Fig.~\ref{fig7} shows the field-dependence of the \tg-symmetry magnon spectrum of \coo\,in the Faraday ($\boldsymbol k \parallel \boldsymbol M \parallel \boldsymbol H$) geometry for different applied pressures at $T=10$ K. Fig.~\subref*{P2a} summarizes the field dependence of the \tg-symmetry magnon energy at $T=10$ K for different applied pressures, and Fig.~\subref*{P2b} shows the amplitude of the \tg-symmetry magnon Raman susceptibility (normalized by the amplitude of the 199 cm$^{-1}$ T$_{2g}$ phonon susceptibility) at $T=10$ K for different applied pressures. The inset of Fig.~\subref*{P2b} summarizes the pressure-dependence of the \tg-symmetry magnon energy of \coo\,for $H=0$ T and $T=10$ K.

\subsection{Discussion and Analysis}
\begin{figure}
\subfloat{\includegraphics[scale=0.25]{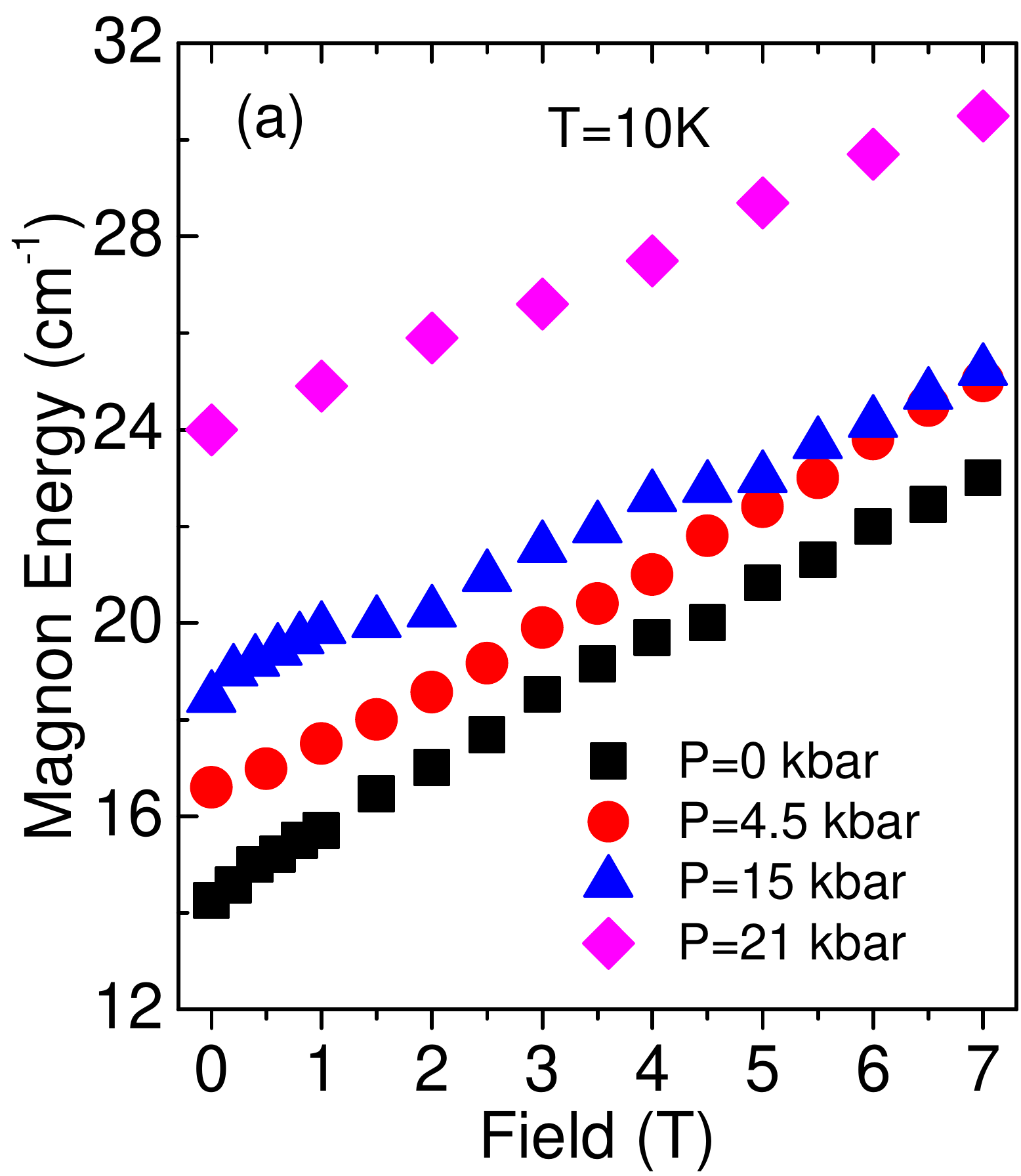}
\label{P2a}}
\quad
\subfloat{\includegraphics[scale=0.25]{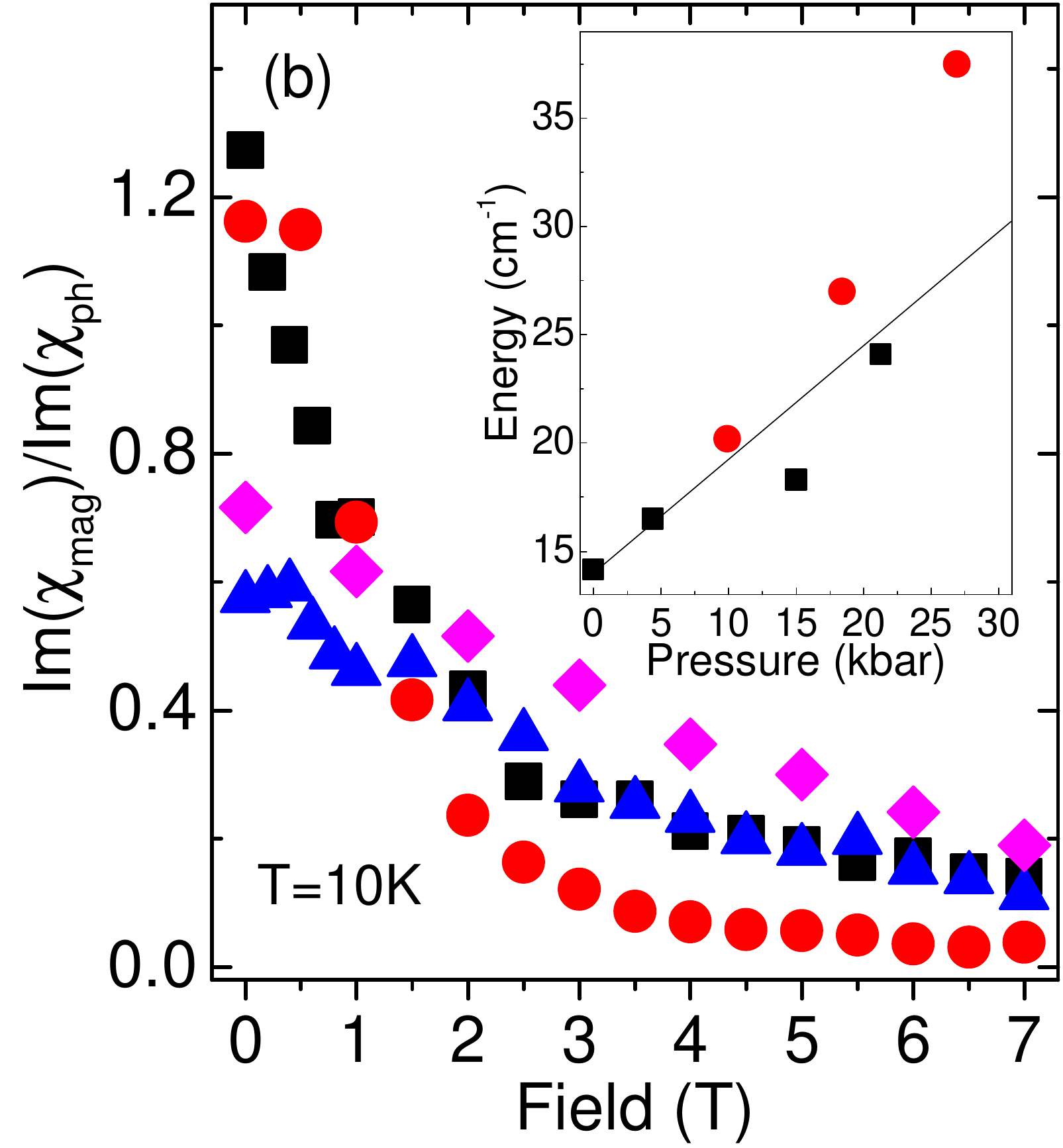}
\label{P2b}}
\caption{\label{fig8}(a) Summary of the field dependences in the Faraday ($\boldsymbol k \parallel \boldsymbol M \parallel \boldsymbol H$) geometry of the \tg-symmetry magnon energy of \coo\,at (filled squares) $T=10$ K and at various pressures, including (filled squares) $P=0$ kbar, (filled circles) $P=4.5$ kbar, (filled triangles) $P=15$ kbar, and (filled diamonds) $P=21$ kbar. (b) Summary of the field dependences of the amplitude of the \tg-symmetry magnon Raman susceptibility normalized to the amplitude of the 199 cm$^{-1}$ T$_{2g}$ phonon Raman susceptibility at $T=10$ K and various pressures, including (filled squares) $P=0$ kbar, (filled circles) $P=4.5$ kbar, (filled triangles) $P=15$ kbar, and (filled diamonds) $P=21$ kbar. ((b) inset) Summary of the pressure-dependence of the \tg-symmetry magnon energy in \coo\,at $T=10$ K and $H=0$ T.}
\end{figure}

The inset of Fig.~\subref*{P2b} shows that the \tg-symmetry magnon energy increases linearly with applied pressure at a rate of $d\omega/dP=$0.46 cm$^{-1}$/kbar. This increase likely reflects a systematic increase in the anisotropy field, $H_A$, with increasing pressure, according to the relationship $\omega\sim(2H_AH_E)^{1/2}$. Additionally, the magnetic field dependence of the Raman spectrum of \coo\,at different fixed pressures summarized in Fig.~\subref*{P2a} shows that the field-dependent slope associated with the \tg-symmetry magnon frequency, $d\omega/dH$, is insensitive to applied pressure up to roughly 21 kbar, indicating that the gyromagnetic ratio associated with Co$^{2+}$ is not strongly affected by these pressures in \coo.

On the other hand, Figs.~\ref{fig7} and \ref{fig8} also show that $H=0$ \tg-symmetry magnon Raman intensity, $\text{Im}\,\chi(\omega)$, systematically decreases relative to the T$_{2g}$ phonon intensity, illustrating that increasing pressure suppresses the magnetic fluctuations and the magneto-optical response in \coo\,by increasing the anisotropy field. Additionally, Fig.~\subref*{P2b} shows that increasing pressure reduces the strong suppression of the \tg-symmetry magnon intensity with increasing magnetic field in the Faraday geometry ($\boldsymbol k \parallel \boldsymbol M \parallel \boldsymbol H$), providing evidence that the magnetodielectric response of \coo\,decreases with increasing pressure. Altogether, these results show that, by tuning magnetic anisotropy and reducing magnetic fluctuations of the Co$^{2+}$ spins, pressure and epitaxial strain can be used as effective tuning parameters for controlling the magnetodielectric response of \coo.

\section{Summary and Conclusions}

In this paper, we showed that the $\boldsymbol q=0$ \tg-symmetry magnon in \coo\,exhibits an anomalously large Raman scattering intensity, which reflects a large magneto-optical response that likely results from large magnetic fluctuations that couple strongly to the dielectric response. The strong suppression of the \tg-symmetry magnon Raman intensity in an applied field is consistent with the magnetodielectric response observed previously in this material, \cite{Muf-2010, Yan-2012} and suggests that the strong magnetodielectric response in \coo\,results from the magnetic-field-induced suppression of magnetic fluctuations that are strongly coupled to phonons. \cite{Law-2003} Using pressure to increase the magnetic anisotropy in \coo, we found that we can suppress the magnetic field-dependence of the \tg-symmetry magnon Raman intensity by suppressing magnetic fluctuations, demonstrating that pressure or epitaxial strain should be an effective means of controlling magnetodielectric behavior and the magneto-optical response in \coo. This Raman study also reveals conditions that are conducive for the substantial magneto-optical responses and magneto-dielectric behaviors in materials, including the presence of strong spin-orbit coupling and weak magnetic anisotropy, both of which create favorable conditions for large magnetic fluctuations that strongly modulate the dielectric response.

\acknowledgements{
Research was supported by the National Science Foundation under Grant NSF DMR 1464090. RDM and DPS thank the Illinois Department of Materials Science and Engineering for support. X-ray diffraction and magnetic susceptibility measurements were performed in the Frederick Seitz Materials Research Laboratory.}

\bibliographystyle{apsrev}

\begin{thebibliography}{58}
\expandafter\ifx\csname natexlab\endcsname\relax\def\natexlab#1{#1}\fi
\expandafter\ifx\csname bibnamefont\endcsname\relax
  \def\bibnamefont#1{#1}\fi
\expandafter\ifx\csname bibfnamefont\endcsname\relax
  \def\bibfnamefont#1{#1}\fi
\expandafter\ifx\csname citenamefont\endcsname\relax
  \def\citenamefont#1{#1}\fi
\expandafter\ifx\csname url\endcsname\relax
  \def\url#1{\texttt{#1}}\fi
\expandafter\ifx\csname urlprefix\endcsname\relax\def\urlprefix{URL }\fi
\providecommand{\bibinfo}[2]{#2}
\providecommand{\eprint}[2][]{\url{#2}}

\bibitem[{\citenamefont{Kimura et~al.}(2003{\natexlab{a}})\citenamefont{Kimura,
  Goto, Shintani, Ishizaka, Arima, and Tokura}}]{Kim-2003a}
\bibinfo{author}{\bibfnamefont{T.}~\bibnamefont{Kimura}},
  \bibinfo{author}{\bibfnamefont{T.}~\bibnamefont{Goto}},
  \bibinfo{author}{\bibfnamefont{H.}~\bibnamefont{Shintani}},
  \bibinfo{author}{\bibfnamefont{K.}~\bibnamefont{Ishizaka}},
  \bibinfo{author}{\bibfnamefont{T.}~\bibnamefont{Arima}}, \bibnamefont{and}
  \bibinfo{author}{\bibfnamefont{Y.}~\bibnamefont{Tokura}},
  \bibinfo{journal}{Nature} \textbf{\bibinfo{volume}{55}}, \bibinfo{pages}{426}
  (\bibinfo{year}{2003}{\natexlab{a}}),
  \urlprefix\url{http://www.nature.com/nature/journal/v426/n6962/full/nature02018.html}.

\bibitem[{\citenamefont{Cheong and Mostovoy}(2007)}]{Che-2007}
\bibinfo{author}{\bibfnamefont{S.-W.} \bibnamefont{Cheong}} \bibnamefont{and}
  \bibinfo{author}{\bibfnamefont{M.}~\bibnamefont{Mostovoy}},
  \bibinfo{journal}{Nat. Mat.} \textbf{\bibinfo{volume}{6}},
  \bibinfo{pages}{13} (\bibinfo{year}{2007}).

\bibitem[{\citenamefont{Bar\'yakhtar and Chupis}(1970)}]{Bar-1970}
\bibinfo{author}{\bibfnamefont{V.~G.} \bibnamefont{Bar\'yakhtar}}
  \bibnamefont{and} \bibinfo{author}{\bibfnamefont{I.~E.}
  \bibnamefont{Chupis}}, \bibinfo{journal}{Sov. Phys. Solid State}
  \textbf{\bibinfo{volume}{11}}, \bibinfo{pages}{2628} (\bibinfo{year}{1970}).

\bibitem[{\citenamefont{Pimenov et~al.}(2006)\citenamefont{Pimenov, Mukhin,
  Ivanov, Travkin, Balbashov, and Loidl}}]{Pim-2006}
\bibinfo{author}{\bibfnamefont{A.}~\bibnamefont{Pimenov}},
  \bibinfo{author}{\bibfnamefont{A.~A.} \bibnamefont{Mukhin}},
  \bibinfo{author}{\bibfnamefont{V.~Y.} \bibnamefont{Ivanov}},
  \bibinfo{author}{\bibfnamefont{V.~D.} \bibnamefont{Travkin}},
  \bibinfo{author}{\bibfnamefont{A.~M.} \bibnamefont{Balbashov}},
  \bibnamefont{and} \bibinfo{author}{\bibfnamefont{A.}~\bibnamefont{Loidl}},
  \bibinfo{journal}{Nature Phys.} \textbf{\bibinfo{volume}{2}},
  \bibinfo{pages}{97} (\bibinfo{year}{2006}).

\bibitem[{\citenamefont{Kida et~al.}(2008{\natexlab{a}})\citenamefont{Kida,
  Ikebe, Takahashi, He, Kaneko, Yamasaki, Shimano, Arima, Nagaosa, and
  Tokura}}]{Kid-2008a}
\bibinfo{author}{\bibfnamefont{N.}~\bibnamefont{Kida}},
  \bibinfo{author}{\bibfnamefont{Y.}~\bibnamefont{Ikebe}},
  \bibinfo{author}{\bibfnamefont{Y.}~\bibnamefont{Takahashi}},
  \bibinfo{author}{\bibfnamefont{J.~P.} \bibnamefont{He}},
  \bibinfo{author}{\bibfnamefont{Y.}~\bibnamefont{Kaneko}},
  \bibinfo{author}{\bibfnamefont{Y.}~\bibnamefont{Yamasaki}},
  \bibinfo{author}{\bibfnamefont{R.}~\bibnamefont{Shimano}},
  \bibinfo{author}{\bibfnamefont{T.}~\bibnamefont{Arima}},
  \bibinfo{author}{\bibfnamefont{N.}~\bibnamefont{Nagaosa}}, \bibnamefont{and}
  \bibinfo{author}{\bibfnamefont{Y.}~\bibnamefont{Tokura}},
  \bibinfo{journal}{Phys. Rev. B} \textbf{\bibinfo{volume}{78}},
  \bibinfo{pages}{104414} (\bibinfo{year}{2008}{\natexlab{a}}),
  \urlprefix\url{http://link.aps.org/doi/10.1103/PhysRevB.78.104414}.

\bibitem[{\citenamefont{Kida et~al.}(2008{\natexlab{b}})\citenamefont{Kida,
  Yamasaki, Shimano, hisa Arima, and Tokura}}]{Kid-2008b}
\bibinfo{author}{\bibfnamefont{N.}~\bibnamefont{Kida}},
  \bibinfo{author}{\bibfnamefont{Y.}~\bibnamefont{Yamasaki}},
  \bibinfo{author}{\bibfnamefont{R.}~\bibnamefont{Shimano}},
  \bibinfo{author}{\bibfnamefont{T.}~\bibnamefont{hisa Arima}},
  \bibnamefont{and} \bibinfo{author}{\bibfnamefont{Y.}~\bibnamefont{Tokura}},
  \bibinfo{journal}{Journal of the Physical Society of Japan}
  \textbf{\bibinfo{volume}{77}}, \bibinfo{pages}{123704}
  (\bibinfo{year}{2008}{\natexlab{b}}),
  \eprint{http://dx.doi.org/10.1143/JPSJ.77.123704},
  \urlprefix\url{http://dx.doi.org/10.1143/JPSJ.77.123704}.

\bibitem[{\citenamefont{Sushkov et~al.}(2008)\citenamefont{Sushkov, Mostovoy,
  Aguilar, Cheong, and Drew}}]{Sus-2008}
\bibinfo{author}{\bibfnamefont{A.~B.} \bibnamefont{Sushkov}},
  \bibinfo{author}{\bibfnamefont{M.}~\bibnamefont{Mostovoy}},
  \bibinfo{author}{\bibfnamefont{R.~V.} \bibnamefont{Aguilar}},
  \bibinfo{author}{\bibfnamefont{S.-W.} \bibnamefont{Cheong}},
  \bibnamefont{and} \bibinfo{author}{\bibfnamefont{H.~D.} \bibnamefont{Drew}},
  \bibinfo{journal}{Journal of Physics: Condensed Matter}
  \textbf{\bibinfo{volume}{20}}, \bibinfo{pages}{434210}
  (\bibinfo{year}{2008}),
  \urlprefix\url{http://stacks.iop.org/0953-8984/20/i=43/a=434210}.

\bibitem[{\citenamefont{Vald\'es~Aguilar
  et~al.}(2009)\citenamefont{Vald\'es~Aguilar, Mostovoy, Sushkov, Zhang, Choi,
  Cheong, and Drew}}]{Val-2009}
\bibinfo{author}{\bibfnamefont{R.}~\bibnamefont{Vald\'es~Aguilar}},
  \bibinfo{author}{\bibfnamefont{M.}~\bibnamefont{Mostovoy}},
  \bibinfo{author}{\bibfnamefont{A.~B.} \bibnamefont{Sushkov}},
  \bibinfo{author}{\bibfnamefont{C.~L.} \bibnamefont{Zhang}},
  \bibinfo{author}{\bibfnamefont{Y.~J.} \bibnamefont{Choi}},
  \bibinfo{author}{\bibfnamefont{S.-W.} \bibnamefont{Cheong}},
  \bibnamefont{and} \bibinfo{author}{\bibfnamefont{H.~D.} \bibnamefont{Drew}},
  \bibinfo{journal}{Phys. Rev. Lett.} \textbf{\bibinfo{volume}{102}},
  \bibinfo{pages}{047203} (\bibinfo{year}{2009}),
  \urlprefix\url{http://link.aps.org/doi/10.1103/PhysRevLett.102.047203}.

\bibitem[{\citenamefont{Stenberg and de~Sousa}(2009)}]{Ste-2009}
\bibinfo{author}{\bibfnamefont{M.~P.~V.} \bibnamefont{Stenberg}}
  \bibnamefont{and} \bibinfo{author}{\bibfnamefont{R.}~\bibnamefont{de~Sousa}},
  \bibinfo{journal}{Phys. Rev. B} \textbf{\bibinfo{volume}{80}},
  \bibinfo{pages}{094419} (\bibinfo{year}{2009}),
  \urlprefix\url{http://link.aps.org/doi/10.1103/PhysRevB.80.094419}.

\bibitem[{\citenamefont{Mochizuki et~al.}(2010)\citenamefont{Mochizuki,
  Furukawa, and Nagaosa}}]{Moc-2010}
\bibinfo{author}{\bibfnamefont{M.}~\bibnamefont{Mochizuki}},
  \bibinfo{author}{\bibfnamefont{N.}~\bibnamefont{Furukawa}}, \bibnamefont{and}
  \bibinfo{author}{\bibfnamefont{N.}~\bibnamefont{Nagaosa}},
  \bibinfo{journal}{Phys. Rev. Lett.} \textbf{\bibinfo{volume}{104}},
  \bibinfo{pages}{177206} (\bibinfo{year}{2010}),
  \urlprefix\url{http://link.aps.org/doi/10.1103/PhysRevLett.104.177206}.

\bibitem[{\citenamefont{Tiwari and Sa}(2010)}]{Tiw-2010}
\bibinfo{author}{\bibfnamefont{S.}~\bibnamefont{Tiwari}} \bibnamefont{and}
  \bibinfo{author}{\bibfnamefont{D.}~\bibnamefont{Sa}},
  \bibinfo{journal}{Journal of Physics: Condensed Matter}
  \textbf{\bibinfo{volume}{22}}, \bibinfo{pages}{225903}
  (\bibinfo{year}{2010}),
  \urlprefix\url{http://stacks.iop.org/0953-8984/22/i=22/a=225903}.

\bibitem[{\citenamefont{Harris}(2011)}]{Har-2011}
\bibinfo{author}{\bibfnamefont{A.~B.} \bibnamefont{Harris}},
  \bibinfo{journal}{arXiv:1011.6672v2 [cond-mat.mtrl-sci]}
  (\bibinfo{year}{2011}), \urlprefix\url{https://arxiv.org/pdf/1011.6672.pdf}.

\bibitem[{\citenamefont{Jones et~al.}(2014)\citenamefont{Jones, Gaw, Doig,
  Prabhakaran, Wheeler, Boothroyd, and Lloyd-Hughes}}]{Jon-2014}
\bibinfo{author}{\bibfnamefont{S.~P.~P.} \bibnamefont{Jones}},
  \bibinfo{author}{\bibfnamefont{S.~M.} \bibnamefont{Gaw}},
  \bibinfo{author}{\bibfnamefont{K.~I.} \bibnamefont{Doig}},
  \bibinfo{author}{\bibfnamefont{D.}~\bibnamefont{Prabhakaran}},
  \bibinfo{author}{\bibfnamefont{E.~M.~H.} \bibnamefont{Wheeler}},
  \bibinfo{author}{\bibfnamefont{A.~T.} \bibnamefont{Boothroyd}},
  \bibnamefont{and}
  \bibinfo{author}{\bibfnamefont{J.}~\bibnamefont{Lloyd-Hughes}},
  \bibinfo{journal}{Nature Communications} \textbf{\bibinfo{volume}{5}},
  \bibinfo{pages}{3787} (\bibinfo{year}{2014}).

\bibitem[{\citenamefont{Cao et~al.}(2015)\citenamefont{Cao, Giustino, and
  Radaelli}}]{Cao-2015}
\bibinfo{author}{\bibfnamefont{K.}~\bibnamefont{Cao}},
  \bibinfo{author}{\bibfnamefont{F.}~\bibnamefont{Giustino}}, \bibnamefont{and}
  \bibinfo{author}{\bibfnamefont{P.~G.} \bibnamefont{Radaelli}},
  \bibinfo{journal}{Phys. Rev. Lett.} \textbf{\bibinfo{volume}{114}},
  \bibinfo{pages}{197201} (\bibinfo{year}{2015}),
  \urlprefix\url{http://link.aps.org/doi/10.1103/PhysRevLett.114.197201}.

\bibitem[{\citenamefont{Kimura et~al.}(2003{\natexlab{b}})\citenamefont{Kimura,
  Kawamoto, Yamada, Azuma, Takano, and Tokura}}]{Kim-2003b}
\bibinfo{author}{\bibfnamefont{T.}~\bibnamefont{Kimura}},
  \bibinfo{author}{\bibfnamefont{S.}~\bibnamefont{Kawamoto}},
  \bibinfo{author}{\bibfnamefont{I.}~\bibnamefont{Yamada}},
  \bibinfo{author}{\bibfnamefont{M.}~\bibnamefont{Azuma}},
  \bibinfo{author}{\bibfnamefont{M.}~\bibnamefont{Takano}}, \bibnamefont{and}
  \bibinfo{author}{\bibfnamefont{Y.}~\bibnamefont{Tokura}},
  \bibinfo{journal}{Phys. Rev. B} \textbf{\bibinfo{volume}{67}},
  \bibinfo{pages}{180401} (\bibinfo{year}{2003}{\natexlab{b}}),
  \urlprefix\url{http://link.aps.org/doi/10.1103/PhysRevB.67.180401}.

\bibitem[{\citenamefont{Lawes et~al.}(2003)\citenamefont{Lawes, Ramirez, Varma,
  and Subramanian}}]{Law-2003}
\bibinfo{author}{\bibfnamefont{G.}~\bibnamefont{Lawes}},
  \bibinfo{author}{\bibfnamefont{A.~P.} \bibnamefont{Ramirez}},
  \bibinfo{author}{\bibfnamefont{C.~M.} \bibnamefont{Varma}}, \bibnamefont{and}
  \bibinfo{author}{\bibfnamefont{M.~A.} \bibnamefont{Subramanian}},
  \bibinfo{journal}{Phys. Rev. Lett.} \textbf{\bibinfo{volume}{91}},
  \bibinfo{pages}{257208} (\bibinfo{year}{2003}),
  \urlprefix\url{http://link.aps.org/doi/10.1103/PhysRevLett.91.257208}.

\bibitem[{\citenamefont{Yang et~al.}(2012)\citenamefont{Yang, Bao, Xue, Zhou,
  Gao, Wang, Wang, Song, Sun, Ren et~al.}}]{Yan-2012}
\bibinfo{author}{\bibfnamefont{S.}~\bibnamefont{Yang}},
  \bibinfo{author}{\bibfnamefont{H.~X.} \bibnamefont{Bao}},
  \bibinfo{author}{\bibfnamefont{D.~Z.} \bibnamefont{Xue}},
  \bibinfo{author}{\bibfnamefont{C.}~\bibnamefont{Zhou}},
  \bibinfo{author}{\bibfnamefont{J.~H.} \bibnamefont{Gao}},
  \bibinfo{author}{\bibfnamefont{Y.}~\bibnamefont{Wang}},
  \bibinfo{author}{\bibfnamefont{J.~Q.} \bibnamefont{Wang}},
  \bibinfo{author}{\bibfnamefont{X.~P.} \bibnamefont{Song}},
  \bibinfo{author}{\bibfnamefont{Z.~B.} \bibnamefont{Sun}},
  \bibinfo{author}{\bibfnamefont{X.~B.} \bibnamefont{Ren}},
  \bibnamefont{et~al.}, \bibinfo{journal}{Journal of Physics D: Applied
  Physics} \textbf{\bibinfo{volume}{45}}, \bibinfo{pages}{265001}
  (\bibinfo{year}{2012}),
  \urlprefix\url{http://stacks.iop.org/0022-3727/45/i=26/a=265001}.

\bibitem[{\citenamefont{Goto et~al.}(2004)\citenamefont{Goto, Kimura, Lawes,
  Ramirez, and Tokura}}]{Got-2004}
\bibinfo{author}{\bibfnamefont{T.}~\bibnamefont{Goto}},
  \bibinfo{author}{\bibfnamefont{T.}~\bibnamefont{Kimura}},
  \bibinfo{author}{\bibfnamefont{G.}~\bibnamefont{Lawes}},
  \bibinfo{author}{\bibfnamefont{A.~P.} \bibnamefont{Ramirez}},
  \bibnamefont{and} \bibinfo{author}{\bibfnamefont{Y.}~\bibnamefont{Tokura}},
  \bibinfo{journal}{Phys. Rev. Lett.} \textbf{\bibinfo{volume}{92}},
  \bibinfo{pages}{257201} (\bibinfo{year}{2004}),
  \urlprefix\url{http://link.aps.org/doi/10.1103/PhysRevLett.92.257201}.

\bibitem[{\citenamefont{Bord\'acs et~al.}(2009)\citenamefont{Bord\'acs, Varjas,
  K\'ezsm\'arki, Mih\'aly, Baldassarre, Abouelsayed, Kuntscher, Ohgushi, and
  Tokura}}]{Bor-2009}
\bibinfo{author}{\bibfnamefont{S.}~\bibnamefont{Bord\'acs}},
  \bibinfo{author}{\bibfnamefont{D.}~\bibnamefont{Varjas}},
  \bibinfo{author}{\bibfnamefont{I.}~\bibnamefont{K\'ezsm\'arki}},
  \bibinfo{author}{\bibfnamefont{G.}~\bibnamefont{Mih\'aly}},
  \bibinfo{author}{\bibfnamefont{L.}~\bibnamefont{Baldassarre}},
  \bibinfo{author}{\bibfnamefont{A.}~\bibnamefont{Abouelsayed}},
  \bibinfo{author}{\bibfnamefont{C.~A.} \bibnamefont{Kuntscher}},
  \bibinfo{author}{\bibfnamefont{K.}~\bibnamefont{Ohgushi}}, \bibnamefont{and}
  \bibinfo{author}{\bibfnamefont{Y.}~\bibnamefont{Tokura}},
  \bibinfo{journal}{Phys. Rev. Lett.} \textbf{\bibinfo{volume}{103}},
  \bibinfo{pages}{077205} (\bibinfo{year}{2009}),
  \urlprefix\url{http://link.aps.org/doi/10.1103/PhysRevLett.103.077205}.

\bibitem[{\citenamefont{Singh et~al.}(2011)\citenamefont{Singh, Maignan, Simon,
  and Martin}}]{Sin-2011}
\bibinfo{author}{\bibfnamefont{K.}~\bibnamefont{Singh}},
  \bibinfo{author}{\bibfnamefont{A.}~\bibnamefont{Maignan}},
  \bibinfo{author}{\bibfnamefont{C.}~\bibnamefont{Simon}}, \bibnamefont{and}
  \bibinfo{author}{\bibfnamefont{C.}~\bibnamefont{Martin}},
  \bibinfo{journal}{Applied Physics Letters} \textbf{\bibinfo{volume}{99}},
  \bibinfo{eid}{172903} (\bibinfo{year}{2011}),
  \urlprefix\url{http://scitation.aip.org/content/aip/journal/apl/99/17/10.1063/1.3656711}.

\bibitem[{\citenamefont{Tomiyasu et~al.}(2004)\citenamefont{Tomiyasu, Fukunaga,
  and Suzuki}}]{Tom-2004}
\bibinfo{author}{\bibfnamefont{K.}~\bibnamefont{Tomiyasu}},
  \bibinfo{author}{\bibfnamefont{J.}~\bibnamefont{Fukunaga}}, \bibnamefont{and}
  \bibinfo{author}{\bibfnamefont{H.}~\bibnamefont{Suzuki}},
  \bibinfo{journal}{Phys. Rev. B} \textbf{\bibinfo{volume}{70}},
  \bibinfo{pages}{214434} (\bibinfo{year}{2004}),
  \urlprefix\url{http://link.aps.org/doi/10.1103/PhysRevB.70.214434}.

\bibitem[{\citenamefont{Tsurkan et~al.}(2013)\citenamefont{Tsurkan, Zherlitsyn,
  Yasin, Felea, Skourski, Deisenhofer, von Nidda, Wosnitza, and
  Loidl}}]{Tsu-2013}
\bibinfo{author}{\bibfnamefont{V.}~\bibnamefont{Tsurkan}},
  \bibinfo{author}{\bibfnamefont{S.}~\bibnamefont{Zherlitsyn}},
  \bibinfo{author}{\bibfnamefont{S.}~\bibnamefont{Yasin}},
  \bibinfo{author}{\bibfnamefont{V.}~\bibnamefont{Felea}},
  \bibinfo{author}{\bibfnamefont{Y.}~\bibnamefont{Skourski}},
  \bibinfo{author}{\bibfnamefont{J.}~\bibnamefont{Deisenhofer}},
  \bibinfo{author}{\bibfnamefont{H.-A.~K.} \bibnamefont{von Nidda}},
  \bibinfo{author}{\bibfnamefont{J.}~\bibnamefont{Wosnitza}}, \bibnamefont{and}
  \bibinfo{author}{\bibfnamefont{A.}~\bibnamefont{Loidl}},
  \bibinfo{journal}{Phys. Rev. Lett.} \textbf{\bibinfo{volume}{110}},
  \bibinfo{pages}{115502} (\bibinfo{year}{2013}),
  \urlprefix\url{http://link.aps.org/doi/10.1103/PhysRevLett.110.115502}.

\bibitem[{\citenamefont{Yamasaki et~al.}(2006)\citenamefont{Yamasaki, Miyasaka,
  Kaneko, He, Arima, and Tokura}}]{Yam-2006}
\bibinfo{author}{\bibfnamefont{Y.}~\bibnamefont{Yamasaki}},
  \bibinfo{author}{\bibfnamefont{S.}~\bibnamefont{Miyasaka}},
  \bibinfo{author}{\bibfnamefont{Y.}~\bibnamefont{Kaneko}},
  \bibinfo{author}{\bibfnamefont{J.-P.} \bibnamefont{He}},
  \bibinfo{author}{\bibfnamefont{T.}~\bibnamefont{Arima}}, \bibnamefont{and}
  \bibinfo{author}{\bibfnamefont{Y.}~\bibnamefont{Tokura}},
  \bibinfo{journal}{Phys. Rev. Lett.} \textbf{\bibinfo{volume}{96}},
  \bibinfo{pages}{207204} (\bibinfo{year}{2006}),
  \urlprefix\url{http://link.aps.org/doi/10.1103/PhysRevLett.96.207204}.

\bibitem[{\citenamefont{Lawes et~al.}(2006)\citenamefont{Lawes, Melot, Page,
  Ederer, Hayward, Proffen, and Seshadri}}]{Law-2006}
\bibinfo{author}{\bibfnamefont{G.}~\bibnamefont{Lawes}},
  \bibinfo{author}{\bibfnamefont{B.}~\bibnamefont{Melot}},
  \bibinfo{author}{\bibfnamefont{K.}~\bibnamefont{Page}},
  \bibinfo{author}{\bibfnamefont{C.}~\bibnamefont{Ederer}},
  \bibinfo{author}{\bibfnamefont{M.~A.} \bibnamefont{Hayward}},
  \bibinfo{author}{\bibfnamefont{T.}~\bibnamefont{Proffen}}, \bibnamefont{and}
  \bibinfo{author}{\bibfnamefont{R.}~\bibnamefont{Seshadri}},
  \bibinfo{journal}{Phys. Rev. B} \textbf{\bibinfo{volume}{74}},
  \bibinfo{pages}{024413} (\bibinfo{year}{2006}),
  \urlprefix\url{http://link.aps.org/doi/10.1103/PhysRevB.74.024413}.

\bibitem[{\citenamefont{Choi et~al.}(2009)\citenamefont{Choi, Okamoto, Huang,
  Chao, Lin, Chen, van Veenendaal, Kaplan, and Cheong}}]{Cho-2009}
\bibinfo{author}{\bibfnamefont{Y.~J.} \bibnamefont{Choi}},
  \bibinfo{author}{\bibfnamefont{J.}~\bibnamefont{Okamoto}},
  \bibinfo{author}{\bibfnamefont{D.~J.} \bibnamefont{Huang}},
  \bibinfo{author}{\bibfnamefont{K.~S.} \bibnamefont{Chao}},
  \bibinfo{author}{\bibfnamefont{H.~J.} \bibnamefont{Lin}},
  \bibinfo{author}{\bibfnamefont{C.~T.} \bibnamefont{Chen}},
  \bibinfo{author}{\bibfnamefont{M.}~\bibnamefont{van Veenendaal}},
  \bibinfo{author}{\bibfnamefont{T.~A.} \bibnamefont{Kaplan}},
  \bibnamefont{and} \bibinfo{author}{\bibfnamefont{S.-W.}
  \bibnamefont{Cheong}}, \bibinfo{journal}{Phys. Rev. Lett.}
  \textbf{\bibinfo{volume}{102}}, \bibinfo{pages}{067601}
  (\bibinfo{year}{2009}),
  \urlprefix\url{http://link.aps.org/doi/10.1103/PhysRevLett.102.067601}.

\bibitem[{\citenamefont{Katsura et~al.}(2005)\citenamefont{Katsura, Nagaosa,
  and Balatsky}}]{Kat-2005}
\bibinfo{author}{\bibfnamefont{H.}~\bibnamefont{Katsura}},
  \bibinfo{author}{\bibfnamefont{N.}~\bibnamefont{Nagaosa}}, \bibnamefont{and}
  \bibinfo{author}{\bibfnamefont{A.~V.} \bibnamefont{Balatsky}},
  \bibinfo{journal}{Phys. Rev. Lett.} \textbf{\bibinfo{volume}{95}},
  \bibinfo{pages}{057205} (\bibinfo{year}{2005}),
  \urlprefix\url{http://link.aps.org/doi/10.1103/PhysRevLett.95.057205}.

\bibitem[{\citenamefont{Mostovoy}(2006)}]{Mos-2006}
\bibinfo{author}{\bibfnamefont{M.}~\bibnamefont{Mostovoy}},
  \bibinfo{journal}{Phys. Rev. Lett.} \textbf{\bibinfo{volume}{96}},
  \bibinfo{pages}{067601} (\bibinfo{year}{2006}),
  \urlprefix\url{http://link.aps.org/doi/10.1103/PhysRevLett.96.067601}.

\bibitem[{\citenamefont{Torgashev et~al.}(2012)\citenamefont{Torgashev,
  Prokhorov, Komandin, Zhukova, Anzin, Talanov, Rabkin, Bush, Dressel, and
  Gorshunov}}]{Tor-2012}
\bibinfo{author}{\bibfnamefont{V.~I.} \bibnamefont{Torgashev}},
  \bibinfo{author}{\bibfnamefont{A.~S.} \bibnamefont{Prokhorov}},
  \bibinfo{author}{\bibfnamefont{G.~A.} \bibnamefont{Komandin}},
  \bibinfo{author}{\bibfnamefont{E.~S.} \bibnamefont{Zhukova}},
  \bibinfo{author}{\bibfnamefont{V.~B.} \bibnamefont{Anzin}},
  \bibinfo{author}{\bibfnamefont{V.~M.} \bibnamefont{Talanov}},
  \bibinfo{author}{\bibfnamefont{L.~M.} \bibnamefont{Rabkin}},
  \bibinfo{author}{\bibfnamefont{A.~A.} \bibnamefont{Bush}},
  \bibinfo{author}{\bibfnamefont{M.}~\bibnamefont{Dressel}}, \bibnamefont{and}
  \bibinfo{author}{\bibfnamefont{B.~P.} \bibnamefont{Gorshunov}},
  \bibinfo{journal}{Physics of the Solid State} \textbf{\bibinfo{volume}{54}},
  \bibinfo{pages}{350} (\bibinfo{year}{2012}), ISSN \bibinfo{issn}{1090-6460},
  \urlprefix\url{http://dx.doi.org/10.1134/S1063783412020321}.

\bibitem[{\citenamefont{Kamenskyi et~al.}(2013)\citenamefont{Kamenskyi,
  Engelkamp, Fischer, Uhlarz, Wosnitza, Gorshunov, Komandin, Prokhorov,
  Dressel, Bush et~al.}}]{Kam-2013}
\bibinfo{author}{\bibfnamefont{D.}~\bibnamefont{Kamenskyi}},
  \bibinfo{author}{\bibfnamefont{H.}~\bibnamefont{Engelkamp}},
  \bibinfo{author}{\bibfnamefont{T.}~\bibnamefont{Fischer}},
  \bibinfo{author}{\bibfnamefont{M.}~\bibnamefont{Uhlarz}},
  \bibinfo{author}{\bibfnamefont{J.}~\bibnamefont{Wosnitza}},
  \bibinfo{author}{\bibfnamefont{B.~P.} \bibnamefont{Gorshunov}},
  \bibinfo{author}{\bibfnamefont{G.~A.} \bibnamefont{Komandin}},
  \bibinfo{author}{\bibfnamefont{A.~S.} \bibnamefont{Prokhorov}},
  \bibinfo{author}{\bibfnamefont{M.}~\bibnamefont{Dressel}},
  \bibinfo{author}{\bibfnamefont{A.~A.} \bibnamefont{Bush}},
  \bibnamefont{et~al.}, \bibinfo{journal}{Phys. Rev. B}
  \textbf{\bibinfo{volume}{87}}, \bibinfo{pages}{134423}
  (\bibinfo{year}{2013}),
  \urlprefix\url{http://link.aps.org/doi/10.1103/PhysRevB.87.134423}.

\bibitem[{\citenamefont{Kushwaha}(2009)}]{Kus-2009}
\bibinfo{author}{\bibfnamefont{A.~K.} \bibnamefont{Kushwaha}},
  \bibinfo{journal}{Chinese Journal of Physics} \textbf{\bibinfo{volume}{47}},
  \bibinfo{pages}{355} (\bibinfo{year}{2009}), ISSN \bibinfo{issn}{0577-9073}.

\bibitem[{\citenamefont{Ptak et~al.}(2014)\citenamefont{Ptak, Mączka, Pikul,
  Tomaszewski, and Hanuza}}]{Pta-2014}
\bibinfo{author}{\bibfnamefont{M.}~\bibnamefont{Ptak}},
  \bibinfo{author}{\bibfnamefont{M.}~\bibnamefont{Mączka}},
  \bibinfo{author}{\bibfnamefont{A.}~\bibnamefont{Pikul}},
  \bibinfo{author}{\bibfnamefont{P.}~\bibnamefont{Tomaszewski}},
  \bibnamefont{and} \bibinfo{author}{\bibfnamefont{J.}~\bibnamefont{Hanuza}},
  \bibinfo{journal}{Journal of Solid State Chemistry}
  \textbf{\bibinfo{volume}{212}}, \bibinfo{pages}{218 } (\bibinfo{year}{2014}),
  ISSN \bibinfo{issn}{0022-4596},
  \urlprefix\url{http://www.sciencedirect.com/science/article/pii/S0022459613005264}.

\bibitem[{\citenamefont{Efthimiopoulos
  et~al.}(2015)\citenamefont{Efthimiopoulos, Liu, Khare, Sarin, Lochbiler,
  Tsurkan, Loidl, Popov, and Wang}}]{Eft-2015}
\bibinfo{author}{\bibfnamefont{I.}~\bibnamefont{Efthimiopoulos}},
  \bibinfo{author}{\bibfnamefont{Z.~T.~Y.} \bibnamefont{Liu}},
  \bibinfo{author}{\bibfnamefont{S.~V.} \bibnamefont{Khare}},
  \bibinfo{author}{\bibfnamefont{P.}~\bibnamefont{Sarin}},
  \bibinfo{author}{\bibfnamefont{T.}~\bibnamefont{Lochbiler}},
  \bibinfo{author}{\bibfnamefont{V.}~\bibnamefont{Tsurkan}},
  \bibinfo{author}{\bibfnamefont{A.}~\bibnamefont{Loidl}},
  \bibinfo{author}{\bibfnamefont{D.}~\bibnamefont{Popov}}, \bibnamefont{and}
  \bibinfo{author}{\bibfnamefont{Y.}~\bibnamefont{Wang}},
  \bibinfo{journal}{Phys. Rev. B} \textbf{\bibinfo{volume}{92}},
  \bibinfo{pages}{064108} (\bibinfo{year}{2015}),
  \urlprefix\url{http://link.aps.org/doi/10.1103/PhysRevB.92.064108}.

\bibitem[{\citenamefont{Kanomata et~al.}(1988)\citenamefont{Kanomata, Tsuda,
  Yasui, and Kaneko}}]{Kan-1988}
\bibinfo{author}{\bibfnamefont{T.}~\bibnamefont{Kanomata}},
  \bibinfo{author}{\bibfnamefont{T.}~\bibnamefont{Tsuda}},
  \bibinfo{author}{\bibfnamefont{H.}~\bibnamefont{Yasui}}, \bibnamefont{and}
  \bibinfo{author}{\bibfnamefont{T.}~\bibnamefont{Kaneko}},
  \bibinfo{journal}{Physics Letters A} \textbf{\bibinfo{volume}{134}},
  \bibinfo{pages}{196 } (\bibinfo{year}{1988}), ISSN \bibinfo{issn}{0375-9601},
  \urlprefix\url{http://www.sciencedirect.com/science/article/pii/0375960188908201}.

\bibitem[{\citenamefont{Tamura}(1993)}]{Tam-1993}
\bibinfo{author}{\bibfnamefont{S.}~\bibnamefont{Tamura}},
  \bibinfo{journal}{Physica B: Condensed Matter}
  \textbf{\bibinfo{volume}{190}}, \bibinfo{pages}{150 } (\bibinfo{year}{1993}),
  ISSN \bibinfo{issn}{0921-4526},
  \urlprefix\url{http://www.sciencedirect.com/science/article/pii/092145269390460N}.

\bibitem[{\citenamefont{Chen et~al.}(2013)\citenamefont{Chen, Yang, Xie, Huang,
  Ling, Zhang, Pi, Sun, and Zhang}}]{Che-2013}
\bibinfo{author}{\bibfnamefont{X.}~\bibnamefont{Chen}},
  \bibinfo{author}{\bibfnamefont{Z.}~\bibnamefont{Yang}},
  \bibinfo{author}{\bibfnamefont{Y.}~\bibnamefont{Xie}},
  \bibinfo{author}{\bibfnamefont{Z.}~\bibnamefont{Huang}},
  \bibinfo{author}{\bibfnamefont{L.}~\bibnamefont{Ling}},
  \bibinfo{author}{\bibfnamefont{S.}~\bibnamefont{Zhang}},
  \bibinfo{author}{\bibfnamefont{L.}~\bibnamefont{Pi}},
  \bibinfo{author}{\bibfnamefont{Y.}~\bibnamefont{Sun}}, \bibnamefont{and}
  \bibinfo{author}{\bibfnamefont{Y.}~\bibnamefont{Zhang}},
  \bibinfo{journal}{Journal of Applied Physics} \textbf{\bibinfo{volume}{113}},
  \bibinfo{eid}{17E129} (\bibinfo{year}{2013}),
  \urlprefix\url{http://scitation.aip.org/content/aip/journal/jap/113/17/10.1063/1.4796172}.

\bibitem[{\citenamefont{Gleason et~al.}(2014)\citenamefont{Gleason, Byrum, Gim,
  Thaler, Abbamonte, MacDougall, Martin, Zhou, and Cooper}}]{Gle-2014}
\bibinfo{author}{\bibfnamefont{S.~L.} \bibnamefont{Gleason}},
  \bibinfo{author}{\bibfnamefont{T.}~\bibnamefont{Byrum}},
  \bibinfo{author}{\bibfnamefont{Y.}~\bibnamefont{Gim}},
  \bibinfo{author}{\bibfnamefont{A.}~\bibnamefont{Thaler}},
  \bibinfo{author}{\bibfnamefont{P.}~\bibnamefont{Abbamonte}},
  \bibinfo{author}{\bibfnamefont{G.~J.} \bibnamefont{MacDougall}},
  \bibinfo{author}{\bibfnamefont{L.~W.} \bibnamefont{Martin}},
  \bibinfo{author}{\bibfnamefont{H.~D.} \bibnamefont{Zhou}}, \bibnamefont{and}
  \bibinfo{author}{\bibfnamefont{S.~L.} \bibnamefont{Cooper}},
  \bibinfo{journal}{Phys. Rev. B} \textbf{\bibinfo{volume}{89}},
  \bibinfo{pages}{134402} (\bibinfo{year}{2014}),
  \urlprefix\url{http://link.aps.org/doi/10.1103/PhysRevB.89.134402}.

\bibitem[{\citenamefont{Gim et~al.}(2016)\citenamefont{Gim, Sethi, Zhao,
  Mitchell, Cao, and Cooper}}]{Gim-2016}
\bibinfo{author}{\bibfnamefont{Y.}~\bibnamefont{Gim}},
  \bibinfo{author}{\bibfnamefont{A.}~\bibnamefont{Sethi}},
  \bibinfo{author}{\bibfnamefont{Q.}~\bibnamefont{Zhao}},
  \bibinfo{author}{\bibfnamefont{J.~F.} \bibnamefont{Mitchell}},
  \bibinfo{author}{\bibfnamefont{G.}~\bibnamefont{Cao}}, \bibnamefont{and}
  \bibinfo{author}{\bibfnamefont{S.~L.} \bibnamefont{Cooper}},
  \bibinfo{journal}{Phys. Rev. B} \textbf{\bibinfo{volume}{93}},
  \bibinfo{pages}{024405} (\bibinfo{year}{2016}),
  \urlprefix\url{http://link.aps.org/doi/10.1103/PhysRevB.93.024405}.

\bibitem[{\citenamefont{Byrum et~al.}(2016)\citenamefont{Byrum, Gleason,
  Thaler, MacDougall, and Cooper}}]{Byr-2016}
\bibinfo{author}{\bibfnamefont{T.}~\bibnamefont{Byrum}},
  \bibinfo{author}{\bibfnamefont{S.~L.} \bibnamefont{Gleason}},
  \bibinfo{author}{\bibfnamefont{A.}~\bibnamefont{Thaler}},
  \bibinfo{author}{\bibfnamefont{G.~J.} \bibnamefont{MacDougall}},
  \bibnamefont{and} \bibinfo{author}{\bibfnamefont{S.~L.}
  \bibnamefont{Cooper}}, \bibinfo{journal}{Phys. Rev. B}
  \textbf{\bibinfo{volume}{93}}, \bibinfo{pages}{184418}
  (\bibinfo{year}{2016}),
  \urlprefix\url{http://link.aps.org/doi/10.1103/PhysRevB.93.184418}.

\bibitem[{\citenamefont{Cazayous et~al.}(2008)\citenamefont{Cazayous, Gallais,
  Sacuto, de~Sousa, Lebeugle, and Colson}}]{Caz-2008}
\bibinfo{author}{\bibfnamefont{M.}~\bibnamefont{Cazayous}},
  \bibinfo{author}{\bibfnamefont{Y.}~\bibnamefont{Gallais}},
  \bibinfo{author}{\bibfnamefont{A.}~\bibnamefont{Sacuto}},
  \bibinfo{author}{\bibfnamefont{R.}~\bibnamefont{de~Sousa}},
  \bibinfo{author}{\bibfnamefont{D.}~\bibnamefont{Lebeugle}}, \bibnamefont{and}
  \bibinfo{author}{\bibfnamefont{D.}~\bibnamefont{Colson}},
  \bibinfo{journal}{Phys. Rev. Lett.} \textbf{\bibinfo{volume}{101}},
  \bibinfo{pages}{037601} (\bibinfo{year}{2008}),
  \urlprefix\url{http://link.aps.org/doi/10.1103/PhysRevLett.101.037601}.

\bibitem[{\citenamefont{Singh et~al.}(2008)\citenamefont{Singh, Katiyar, and
  Scott}}]{Sin-2008}
\bibinfo{author}{\bibfnamefont{M.~K.} \bibnamefont{Singh}},
  \bibinfo{author}{\bibfnamefont{R.~S.} \bibnamefont{Katiyar}},
  \bibnamefont{and} \bibinfo{author}{\bibfnamefont{J.~F.} \bibnamefont{Scott}},
  \bibinfo{journal}{Journal of Physics: Condensed Matter}
  \textbf{\bibinfo{volume}{20}}, \bibinfo{pages}{252203}
  (\bibinfo{year}{2008}),
  \urlprefix\url{http://stacks.iop.org/0953-8984/20/i=25/a=252203}.

\bibitem[{\citenamefont{Rovillain et~al.}(2010)\citenamefont{Rovillain,
  Cazayous, Gallais, Sacuto, Measson, and Sakata}}]{Rov-2010}
\bibinfo{author}{\bibfnamefont{P.}~\bibnamefont{Rovillain}},
  \bibinfo{author}{\bibfnamefont{M.}~\bibnamefont{Cazayous}},
  \bibinfo{author}{\bibfnamefont{Y.}~\bibnamefont{Gallais}},
  \bibinfo{author}{\bibfnamefont{A.}~\bibnamefont{Sacuto}},
  \bibinfo{author}{\bibfnamefont{M.-A.} \bibnamefont{Measson}},
  \bibnamefont{and} \bibinfo{author}{\bibfnamefont{H.}~\bibnamefont{Sakata}},
  \bibinfo{journal}{Phys. Rev. B} \textbf{\bibinfo{volume}{81}},
  \bibinfo{pages}{054428} (\bibinfo{year}{2010}),
  \urlprefix\url{http://link.aps.org/doi/10.1103/PhysRevB.81.054428}.

\bibitem[{\citenamefont{Rovillain et~al.}(2011)\citenamefont{Rovillain,
  Cazayous, Gallais, Measson, Sacuto, Sakata, and Mochizuki}}]{Rov-2011}
\bibinfo{author}{\bibfnamefont{P.}~\bibnamefont{Rovillain}},
  \bibinfo{author}{\bibfnamefont{M.}~\bibnamefont{Cazayous}},
  \bibinfo{author}{\bibfnamefont{Y.}~\bibnamefont{Gallais}},
  \bibinfo{author}{\bibfnamefont{M.-A.} \bibnamefont{Measson}},
  \bibinfo{author}{\bibfnamefont{A.}~\bibnamefont{Sacuto}},
  \bibinfo{author}{\bibfnamefont{H.}~\bibnamefont{Sakata}}, \bibnamefont{and}
  \bibinfo{author}{\bibfnamefont{M.}~\bibnamefont{Mochizuki}},
  \bibinfo{journal}{Phys. Rev. Lett.} \textbf{\bibinfo{volume}{107}},
  \bibinfo{pages}{027202} (\bibinfo{year}{2011}),
  \urlprefix\url{http://link.aps.org/doi/10.1103/PhysRevLett.107.027202}.

\bibitem[{\citenamefont{Mufti et~al.}(2010)\citenamefont{Mufti, Nugroho, Blake,
  and Palstra}}]{Muf-2010}
\bibinfo{author}{\bibfnamefont{N.}~\bibnamefont{Mufti}},
  \bibinfo{author}{\bibfnamefont{A.~A.} \bibnamefont{Nugroho}},
  \bibinfo{author}{\bibfnamefont{G.~R.} \bibnamefont{Blake}}, \bibnamefont{and}
  \bibinfo{author}{\bibfnamefont{T.~T.~M.} \bibnamefont{Palstra}},
  \bibinfo{journal}{Journal of Physics: Condensed Matter}
  \textbf{\bibinfo{volume}{22}}, \bibinfo{pages}{075902}
  (\bibinfo{year}{2010}),
  \urlprefix\url{http://stacks.iop.org/0953-8984/22/i=7/a=075902}.

\bibitem[{\citenamefont{Ohgushi et~al.}(2008)\citenamefont{Ohgushi, Okimoto,
  Ogasawara, Miyasaka, and Tokura}}]{Ohg-2008}
\bibinfo{author}{\bibfnamefont{K.}~\bibnamefont{Ohgushi}},
  \bibinfo{author}{\bibfnamefont{Y.}~\bibnamefont{Okimoto}},
  \bibinfo{author}{\bibfnamefont{T.}~\bibnamefont{Ogasawara}},
  \bibinfo{author}{\bibfnamefont{S.}~\bibnamefont{Miyasaka}}, \bibnamefont{and}
  \bibinfo{author}{\bibfnamefont{Y.}~\bibnamefont{Tokura}},
  \bibinfo{journal}{Journal of the Physical Society of Japan}
  \textbf{\bibinfo{volume}{77}}, \bibinfo{pages}{034713}
  (\bibinfo{year}{2008}), \eprint{http://dx.doi.org/10.1143/JPSJ.77.034713},
  \urlprefix\url{http://dx.doi.org/10.1143/JPSJ.77.034713}.

\bibitem[{\citenamefont{B\'{e}rar and Baldinozzi}(1998)}]{xnd}
\bibinfo{author}{\bibfnamefont{J.}~\bibnamefont{B\'{e}rar}} \bibnamefont{and}
  \bibinfo{author}{\bibfnamefont{G.}~\bibnamefont{Baldinozzi}},
  \bibinfo{journal}{IUCr-CPD Newsletter} \textbf{\bibinfo{volume}{20}},
  \bibinfo{pages}{3} (\bibinfo{year}{1998}).

\bibitem[{\citenamefont{Kim et~al.}(2011)\citenamefont{Kim, Chen, Wang, Nelson,
  Budakian, Abbamonte, and Cooper}}]{Kim-2011}
\bibinfo{author}{\bibfnamefont{M.}~\bibnamefont{Kim}},
  \bibinfo{author}{\bibfnamefont{X.~M.} \bibnamefont{Chen}},
  \bibinfo{author}{\bibfnamefont{X.}~\bibnamefont{Wang}},
  \bibinfo{author}{\bibfnamefont{C.~S.} \bibnamefont{Nelson}},
  \bibinfo{author}{\bibfnamefont{R.}~\bibnamefont{Budakian}},
  \bibinfo{author}{\bibfnamefont{P.}~\bibnamefont{Abbamonte}},
  \bibnamefont{and} \bibinfo{author}{\bibfnamefont{S.~L.}
  \bibnamefont{Cooper}}, \bibinfo{journal}{Phys. Rev. B}
  \textbf{\bibinfo{volume}{84}}, \bibinfo{pages}{174424}
  (\bibinfo{year}{2011}),
  \urlprefix\url{http://link.aps.org/doi/10.1103/PhysRevB.84.174424}.

\bibitem[{\citenamefont{Cottam and Lockwood}(1986)}]{Cot-1986}
\bibinfo{author}{\bibfnamefont{M.~G.} \bibnamefont{Cottam}} \bibnamefont{and}
  \bibinfo{author}{\bibfnamefont{D.~J.} \bibnamefont{Lockwood}},
  \emph{\bibinfo{title}{{Light Scattering in Magnetic Solids}}}
  (\bibinfo{publisher}{Wiley-Interscience}, \bibinfo{address}{New York},
  \bibinfo{year}{1986}).

\bibitem[{\citenamefont{Ramdas and Rodriguez}(1991)}]{Ram-1991}
\bibinfo{author}{\bibfnamefont{A.~K.} \bibnamefont{Ramdas}} \bibnamefont{and}
  \bibinfo{author}{\bibfnamefont{S.}~\bibnamefont{Rodriguez}},
  \bibinfo{journal}{Topics in Applied Physics} \textbf{\bibinfo{volume}{68}},
  \bibinfo{pages}{137} (\bibinfo{year}{1991}).

\bibitem[{\citenamefont{Kaplan and Kittel}(1953)}]{Kap-1953}
\bibinfo{author}{\bibfnamefont{J.}~\bibnamefont{Kaplan}} \bibnamefont{and}
  \bibinfo{author}{\bibfnamefont{C.}~\bibnamefont{Kittel}},
  \bibinfo{journal}{J. Chem. Phys.} \textbf{\bibinfo{volume}{5}},
  \bibinfo{pages}{760} (\bibinfo{year}{1953}),
  \urlprefix\url{http://scitation.aip.org/content/aip/journal/jcp/21/4/10.1063/1.1699018}.

\bibitem[{\citenamefont{Brinkman and Elliott}(1966)}]{Bri-1966}
\bibinfo{author}{\bibfnamefont{W.~F.} \bibnamefont{Brinkman}} \bibnamefont{and}
  \bibinfo{author}{\bibfnamefont{R.~J.} \bibnamefont{Elliott}},
  \bibinfo{journal}{Proceedings of the Royal Society of London, Series A:
  Mathematical, Physical and Engineering Sciences}
  \textbf{\bibinfo{volume}{294}}, \bibinfo{pages}{343} (\bibinfo{year}{1966}),
  ISSN \bibinfo{issn}{0080-4630},
  \eprint{http://rspa.royalsocietypublishing.org/content/294/1438/343.full.pdf},
  \urlprefix\url{http://rspa.royalsocietypublishing.org/content/294/1438/343}.

\bibitem[{\citenamefont{Sahni and Venkataraman}(1974)}]{Sah-1974}
\bibinfo{author}{\bibfnamefont{V.}~\bibnamefont{Sahni}} \bibnamefont{and}
  \bibinfo{author}{\bibfnamefont{G.}~\bibnamefont{Venkataraman}},
  \bibinfo{journal}{Advances in Physics} \textbf{\bibinfo{volume}{23}},
  \bibinfo{pages}{547} (\bibinfo{year}{1974}),
  \eprint{http://dx.doi.org/10.1080/00018737400101391},
  \urlprefix\url{http://dx.doi.org/10.1080/00018737400101391}.

\bibitem[{\citenamefont{Demokritov et~al.}(1987)\citenamefont{Demokritov,
  Kreines, and Kudinov}}]{Dem-1987}
\bibinfo{author}{\bibfnamefont{S.~O.} \bibnamefont{Demokritov}},
  \bibinfo{author}{\bibfnamefont{N.~M.} \bibnamefont{Kreines}},
  \bibnamefont{and} \bibinfo{author}{\bibfnamefont{V.~I.}
  \bibnamefont{Kudinov}}, \bibinfo{journal}{Sov. Phys. JETP}
  \textbf{\bibinfo{volume}{65}}, \bibinfo{pages}{389} (\bibinfo{year}{1987}).

\bibitem[{\citenamefont{Kumar et~al.}(2011)\citenamefont{Kumar, Scott, and
  Katiyar}}]{Kum-2011}
\bibinfo{author}{\bibfnamefont{A.}~\bibnamefont{Kumar}},
  \bibinfo{author}{\bibfnamefont{J.~F.} \bibnamefont{Scott}}, \bibnamefont{and}
  \bibinfo{author}{\bibfnamefont{R.~S.} \bibnamefont{Katiyar}},
  \bibinfo{journal}{Applied Physics Letters} \textbf{\bibinfo{volume}{99}},
  \bibinfo{eid}{062504} (\bibinfo{year}{2011}),
  \urlprefix\url{http://scitation.aip.org/content/aip/journal/apl/99/6/10.1063/1.3624845}.

\bibitem[{\citenamefont{Bar\'yakhtar et~al.}(1983)\citenamefont{Bar\'yakhtar,
  Pashkevich, and Sobolev}}]{Bar-1983}
\bibinfo{author}{\bibfnamefont{V.~G.} \bibnamefont{Bar\'yakhtar}},
  \bibinfo{author}{\bibfnamefont{Y.~G.} \bibnamefont{Pashkevich}},
  \bibnamefont{and} \bibinfo{author}{\bibfnamefont{V.~L.}
  \bibnamefont{Sobolev}}, \bibinfo{journal}{Sov. Phys. JETP}
  \textbf{\bibinfo{volume}{58}}, \bibinfo{pages}{945} (\bibinfo{year}{1983}).

\bibitem[{\citenamefont{Borovik-Romanov and Kreines}(1988)}]{Bor-1988}
\bibinfo{author}{\bibfnamefont{A.~S.} \bibnamefont{Borovik-Romanov}}
  \bibnamefont{and} \bibinfo{author}{\bibfnamefont{N.~M.}
  \bibnamefont{Kreines}}, in \emph{\bibinfo{booktitle}{{Spin Waves and Magnetic
  Excitations}}} (\bibinfo{publisher}{North-Holland},
  \bibinfo{address}{Amsterdam}, \bibinfo{year}{1988}),
  chap.~\bibinfo{chapter}{2}.

\bibitem[{\citenamefont{Vitebskii et~al.}(1991)\citenamefont{Vitebskii,
  Yeremenko, Pashkevich, Sobolev, and Fedorov}}]{Vit-1991}
\bibinfo{author}{\bibfnamefont{I.}~\bibnamefont{Vitebskii}},
  \bibinfo{author}{\bibfnamefont{A.}~\bibnamefont{Yeremenko}},
  \bibinfo{author}{\bibfnamefont{Y.}~\bibnamefont{Pashkevich}},
  \bibinfo{author}{\bibfnamefont{V.}~\bibnamefont{Sobolev}}, \bibnamefont{and}
  \bibinfo{author}{\bibfnamefont{S.}~\bibnamefont{Fedorov}},
  \bibinfo{journal}{Physica C: Superconductivity}
  \textbf{\bibinfo{volume}{178}}, \bibinfo{pages}{189 } (\bibinfo{year}{1991}),
  ISSN \bibinfo{issn}{0921-4534},
  \urlprefix\url{http://www.sciencedirect.com/science/article/pii/092145349190175X}.

\bibitem[{\citenamefont{Altshuler and Kozyrev}(1974)}]{Alt-1974}
\bibinfo{author}{\bibfnamefont{S.~A.} \bibnamefont{Altshuler}}
  \bibnamefont{and} \bibinfo{author}{\bibfnamefont{B.~M.}
  \bibnamefont{Kozyrev}}, \emph{\bibinfo{title}{{Electron Paramagnetic
  Resonance in Compounds of TransitionElements}}} (\bibinfo{publisher}{Wiley},
  \bibinfo{address}{New York}, \bibinfo{year}{1974}).

\bibitem[{\citenamefont{Smolenskiĭ and Chupis}(1982)}]{Smo-1982}
\bibinfo{author}{\bibfnamefont{G.~A.} \bibnamefont{Smolenskiĭ}}
  \bibnamefont{and} \bibinfo{author}{\bibfnamefont{I.~E.}
  \bibnamefont{Chupis}}, \bibinfo{journal}{Soviet Physics Uspekhi}
  \textbf{\bibinfo{volume}{25}}, \bibinfo{pages}{475} (\bibinfo{year}{1982}),
  \urlprefix\url{http://stacks.iop.org/0038-5670/25/i=7/a=R02}.

\end{thebibliography}


\end{document}